\documentclass{aa}
\usepackage{graphics}

\begin{document}

\newcommand{\fe}{\ion{Fe}{II}}
\newcommand{\h}{H$_2$}
\newcommand{\kms}{km\,s$^{-1}$ }
\newcommand{\um}{$\mu$m}
\newcommand{\JJ}{{\it J}}
\newcommand{\HH}{{\it H}}
\newcommand{\KK}{{\it K}}
\newcommand{\LL}{{\it L}}
\newcommand{\mdot}{\dot{M}}
\newcommand{\rk}{$r_{K}$}
\newcommand{\rh}{$r_{H}$}
\newcommand{\Av}{$A_{\rm V}$}

\title{Probing the embedded YSOs of the R CrA region through VLT-ISAAC spectroscopy
\thanks{Based on observations collected at the European Southern 
Observatory, Chile (ESO N.069.C-0269, 070.C.-0138, 063.I-0691)}}
\author{B. Nisini\inst{1}, S. Antoniucci\inst{1,2}, T. Giannini\inst{1}, 
D. Lorenzetti\inst{1} }
\offprints{Brunella Nisini, nisini@mporzio.astro.it}

\institute{INAF-Osservatorio Astronomico di Roma, Via di Frascati 33, I-00040 Monteporzio Catone,
Italy \and Universit\`a degli Studi `Tor Vergata', via della Ricerca Scientifica
1, 00133 Roma, Italy
}
%
%
\date{Received date; Accepted date}
%
%
%
\titlerunning{Embedded sources in R CrA}
\authorrunning{B. Nisini et al.}

\abstract{Near IR spectra at low (R$\sim$800) and medium (R$\sim$9000) 
resolution, obtained with ISAAC at VLT, have been used to pose 
constraints on the evolutionary state and accretion properties of a 
sample of five embedded YSOs located in the R CrA core. This sample includes
three Class I sources (HH100 IR, IRS2 and IRS5), and two sources with
NIR excesses (IRS6 and IRS3). IRS5 and IRS6 have been discovered to be binaries
with a separation between the two components of 97 and 78 AU, respectively.
Absorption lines, typical of late-type photospheres, have been detected
in the medium resolution spectra of all the observed targets,
including HH100 IR and IRS2 which have high values of infrared
continuum veiling ($r_k$ = 6 and 3, respectively). These two sources
also present low resolution spectra rich of emission lines (HI, CO 
and plenty of other permitted lines from neutral atoms) likely originating 
in the disk-star-wind connected regions. 
Among the features observed in HH100 IR and IRS2, Na I at 2.205 \um\, and CO at 2.3\um,
which are more commonly used for stellar classification, 
are detected in emission instead of absorption. 
Several strong photospheric lines, which lie around 2.12 and 2.23\um\, 
and whose ratio is sensitive to both effective temperature and gravity,       
are proposed as independent diagnostic tools for this type of sources.
We derived spectral types, veiling and stellar luminosity 
of the five observed sources, which in turn have been used to infer their 
mass (ranging between 0.3-1.2 M$_\odot$) and age (between 10$^{5}$ and 10$^{6}$ yr)
adopting pre-main sequence evolutionary tracks.
We find that in HH100 IR and IRS2 most of the bolometric luminosity is due 
to accretion ($L_{acc}$/$L_{bol}$ $\sim$0.8 and 0.6
respectively), while the other three investigated sources , including the Class I
object IRS5a, present a low accretion activity ($L_{acc}$/$L_{bol} <$ 0.2).
Mass accretion rates of the order of 2\,10$^{-6}$
and 3\,10$^{-7}$ M$_\odot$\,yr$^{-1}$ are derived for HH100 IR and IRS2,
respectively, i.e. higher by an order of magnitude with respect to the average values 
derived in T Tauri stars. 
We observe a general correlation between the 
accretion luminosity, the IR veiling and the emission line activity of 
the sources. In particular, we find that the correlation between $L_{acc}$ and
$L_{Br\gamma}$, previously reported for optical T Tauri stars, can be extended also to the
embedded sources, up to at least one order of magnitude larger line luminosity.
A correlation between the accretion activity and the 
spectral energy distribution slope is recognizable but with the notable exception 
of IRS5a. Our analysis therefore shows how the definition of the evolutionary stage 
of deeply embedded YSOs by means of IR colors needs to be more carefully
refined.

}
\maketitle

\section{Introduction}

During the first evolutionary stages of a young star, the
protostellar system is characterized by a variety of phenomena occurring in different
regions of the circumstellar environment. These include the accretion disk,
the magnetospheric accretion region, collimated jets and, in more embedded
sources, the still not dispersed dusty envelopes.
In T Tauri systems, where the contribution from the dusty envelope has become
negligible, there is the possibility to disentangle, through observations at
different wavelengths, the emission from the star itself from that
of the different active circumstellar regions, and to define in detail the
properties of these latters. Hence the stellar parameters of T Tauri stars
have been derived through the observations of their optical and IR 
photospheric features (e.g.Luhman \& Rieke 1998), their accretion region
revealed through the observation of the UV excesses over the stellar photosphere
(e.g. Gullbring et al. 1998), and finally their collimated jets probed 
down to close to their base through high resolution observations of optical 
forbidden lines (e.g. Bacciotti et al. 2000). The possibility to separate
the different emission components contributing to the stellar system has
greatly advanced the comparison of observations with models
describing the accretion and ejection mechanisms in these stars, as well
as the disk evolution and dispersal.
It is however fundamental to understand how and if these models works for less evolved
systems, where the circumstellar activity is much higher and the accretion
luminosity is expected to dominate over the luminosity of the still forming star.
To this aim, a sample of interesting objects to be investigated are 
sources with no optical counterpart, characterized by steeply rising IR SEDs
and with infrared colors indicating the presence of intrinsic excesses due
to circumstellar dust (the so-called Class I sources). 
The standard paradigm is that most of these sources are protostars still 
embedded in their original infalling envelope and expected to derive most of their luminosity 
from accretion through a massive circumstellar disk.
For such more embedded sources, however, the large extinction coupled with strong
emission excesses due to the circumstellar envelope, make the investigation 
of phenomena occurring very close to the protostar extremely difficult and thus
their real nature remains often obscure.
Infrared spectroscopy in this respect can significantly improve the 
study of the close environments of these sources, since different gas 
features which are present in the near IR atmospheric windows can be related to 
different emitting regions. The near IR 
spectra of these objects are often characterized by the presence of 
ionized (e.g. Fe$^+$) , neutral (e.g. HI, He) and molecular (CO, H$_2$) emission lines.
Such lines originate in the inner and warm regions of the disk, in the 
accretion region connecting the disk to the star, and in the related 
phenomenon of winds and jets, and consequently they have been used to probe
such different environments (e.g. Nisini et al. 2002, Najita et al. 1996,  Davis et al.
2002, Muzerolle et al. 1998).
The large values of the IR excess characteristic of these source, however, 
produce a large veiling which makes  
extremely difficult to infer their stellar properties by means
of their underlying photospheric features and therefore to correlate the 
spectral probes of circumstellar activity with the physical parameters of the
star itself.

Recently, the first direct observation of absorption lines 
from the photosphere of a Class I source in the $\rho$ Oph cloud has been reported by 
Greene \& Lada (2002). The observed features are typical of late type stars, 
confirming the low mass nature of these sources, and have been used to determine
for the first time the stellar properties of one of
such highly obscured objects, which resulted to have $\sim$ 70\% of its
luminosity due to accretion.
The potentiality offered by high-resolution near IR spectroscopy to
reveal the central protostellar source characteristics 
opens in turn the possibility to both derive the mass and ages of these
young objects, as well as infer quantitatively the mass accretion rate and
accretion luminosity to be compared with those of expected more evolved 
sources. 
In order to make this comparison significant it is also important
to have this information in a sample as large as possible of embedded sources, in such
a way to better constrain their global accretion properties.
In this paper we report the results of an analysis of medium/high resolution near 
IR spectra of a sample of embedded young stellar objects of the nearby R CrA 
star forming cloud (D=130 pc, Marraco \& Rydgren 1981).
The selection of sources from the same cloud assures that 
the properties of the sources reflect intrinsic differences and not the different
characteristics of the hosting cloud. The aim of this paper is twofold: 1) to infer,
through the observations of photospheric features, the stellar
physical parameters of the considered objects and
from these indirectly derive their age,  accretion luminosity and mass accretion rates.
This information will be used to understand to which extent the spectral energy
distribution of embedded IR sources is really indicative of their evolutionary stage;
2) to establish if there is a correlation between the accretion luminosity of the 
embedded stars and the occurrence and luminosity of their IR emission lines in
order to have an indirect instrument to quantitatively derive 
the accretion rate from simple observational tools. 
We describe the considered sample in \S 2 and the observations 
obtained at VLT with the ISAAC instrument in \S 3. In \S 4 and \S 5 the
characteristics of the emission line spectra are described, while in \S 6 and 7
we present the procedure adopted to derive the stellar and circumstellar 
properties from the detected absorption features. 
Finally, in \S 8 we discuss the implications of our findings. 

\section{Description of the selected sample}

The sources investigated in this paper have been taken from the
sample of IR young stellar objects in the R CrA cloud core listed by
Wilking et al. (1997). From this list,  
we
have selected all the IR sources with \KK\, magnitudes smaller than 10.5 mag 
and IR colors compatible with the locus of young embedded protostars with intrinsic IR excesses 
(Lada \& Adams 1992). 
The list of sources is presented in Table 1 with their coordinates,
IR magnitudes and luminosities, when available. All the sources have luminosities ranging 
between 1 and 15 L$_{\odot}$, and thus they are expected to be low mass young
stellar objects.\\
Three of the considered sources, namely HH100 IR, IRS2 and IRS5, can be classified as 
Class I protostars  on the basis of their spectral energy distribution
(having spectral index between 2 and 10$\mu$m $>$ 0, Wilking et al. 1986). 
These three sources are known to have large dusty envelopes as testified by the
presence of strong ice bands and silicate features in their 3, 5 and 10$\mu$m spectra 
(Brooke et al. 1999, Whittet et al. 1996, Alexander et al. 2003, Chen \& Graham 1993,
Pontoppidan et al. 2003). HH100 IR has been also suggested as the driving source
of the Herbig Haro objects HH101 and HH99 (Hartigan \& Graham 1987)
and of a bipolar molecular outflow (Anglada et al. 1989). 

\begin{table*}
\label{tab:sample}
\caption[]{Observed sample}
\vspace{0.5cm}
    \begin{tabular}[h]{cccccccccc}
      \hline \\[-5pt]
Source  & $\alpha$ (2000)&  $\delta$ (2000) & \KK & (\KK-\LL)&(\HH-\KK)& (\JJ-\HH)& $\alpha^{a}$ & Ref.$^{b}$ 
& L$_{bol}$\\
       &                 &                  & \multicolumn{4}{c}{magnitudes} & & & (L$_{\sun}$)\\[+5pt]
      \hline \\[-5pt]
IRS2 (TS13.1)& 19 01 41.5 & $-$36 58 29    & 7.16 & 2.72 & 2.73 & 4.07 & 0.98 & TS &12$^c$\\
                              & &          & 7.2  & --   & 2.7  & 3.9  &      &SOFI&\\
IRS5 (TS2.4) & 19 01 48.0 & $-$36 57 19    & 10.33& 2.75 & 3.32 & 3.89 & 1.12 &TS& 3$^c$\\
IRS5a                         & &          & 10.9 &  --  & 3.8  & --   &      &ISAAC&2\\
IRS5b                         & &          & 11.5 &  --  & 3.4  & --   &      &ISAAC&\\
IRS6 (TS2.3) & 19 01 50.6 & $-$36 56 38    & 10.34& 1.74 & 2.37 & 3.64 & $\sim$0 &TS& $\sim$0.6$^d$\\
IRS6a                        & &           & 11.1 & --   & 2.2  & --   &      &ISAAC&$\sim$0.3\\
IRS6b                         & &          & 11.7 & --   & 3.1  & --   &      &ISAAC&\\
HH100 IR (TS2.6)& 19 01 50.7 & $-$36 58 10 &  8.04& 3.36 & 3.5  & 4.87 & 1.47 &TS& 14$^c$\\
                     & &                   &  7.4 & --   & 2.5  & 4.1  &      &SOFI&\\
IRS3 (TS4.4)& 19 02 05.1 & $-$36 58 56     &  8.64& --   & 2.02 & 2.26 &      &TS& $\sim$0.3$^d$\\
      & &                                  &  9.50& --   & 0.93 & 0.14 &      &B92&\\
\hline \\[+5pt]
      \end{tabular}

~$^{a}$Spectral index ($d\,log(\lambda\,F_{\lambda})/d\,log\lambda$) between 2 and
10$\mu$m.\\
~$^{b}$Origin of photometric data: TS = Taylor \& Storey (1984), SOFI = our NTT-SOFI
observations (March 2003), ISAAC = VLT-ISAAC archive data (1999), B92 = Burton et al. (1992). \\
~$^{c}$From Wilking et al. 1992\\
~$^{d}$estimated from the photometry up to the ${\it Q}$ band given by Wilking et al. 1986\\
\end{table*}

\section{Observations}

\begin{figure}[!ht]
\resizebox{\hsize}{!}{\includegraphics{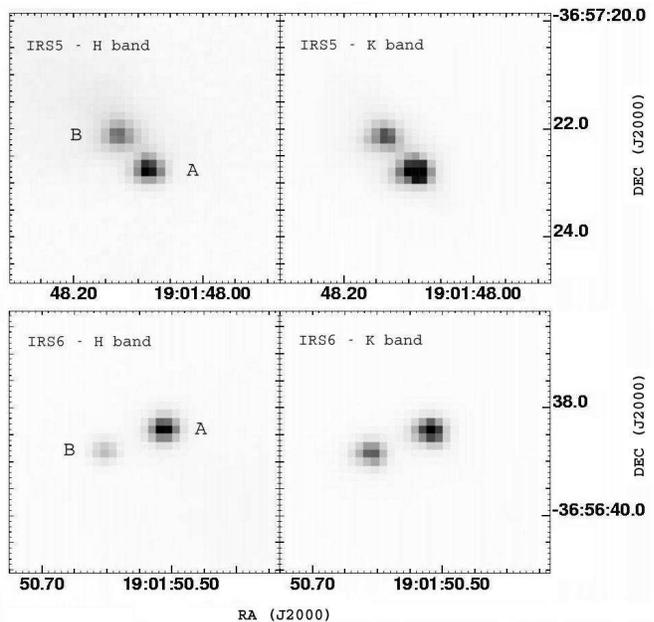}}
   \caption{\label{fig:binary}VLT-ISAAC \HH\, and \KK\, images of the sources 
   IRS5 and IRS6 were the discovered binary components {\it a} and {\it b}
   are indicated.}
\end{figure}

Observations were carried out at VLT/UT1 with the ISAAC spectrograph in SW
mode on 12 and 13 July, 2002. 
At the spatial resolution provided by ISAAC (i.e. 0.14 arcsec/pixel), the sources 
IRS5 and IRS6 were recognized to be binaries, with a separation 
of $\sim$0.75 and 0.6 arcsec (i.e. 97 and 78 AU), respectively. 
Figure \ref{fig:binary} shows \HH\, and \KK\,
images obtained from VLT pre-imaging of these stars, in which the two 
components that we named {\it a} and {\it b} are clearly visible. Due to their alignment
in the sky, a single slit, with a position 
angle of 32.6$^{\circ}$,
was adopted to observe simultaneously IRS5a and b, and IRS6a.

Low resolution spectra through the \JJ (1.1-1.4$\mu$m),
 {\it SH} (1.4-1.82$\mu$m) and {\it SK} (1.82-2.5$\mu$m) filters
 were acquired with a slit aperture of 0\farcs6 corresponding to
 a resolution from 750 (in the \KK\, band) to 960 (in the \JJ\, band).
For each source, medium resolution spectra were also acquired in two
spectral segments centered at 2.261\um\, and 2.161\um\, each
covering $\sim$0.122\um\, around the central wavelength.
HH100 IR, IRS2 and IRS3 were also observed in an additional segment in the \HH\,
band, centered at 1.6295\um\, and covering $\sim$0.080\um.  For these observations
we used the 0\farcs3 slit correspondig to a nominal resolution of
 R$\sim$8000 and R$\sim$9000 for the \KK\, and \HH\, observations, respectively.
Spectra of standard stars of O spectral type were also obtained in both low and
medium resolution at an airmass similar to the
scientific spectra to correct for telluric absorption. The telluric spectra 
were carefully cleaned by any intrinsic HI and He absorption feature before being used.
In the low resolution spectra, wavelength calibrations were performed 
using a Xenon lamp spectrum taken at the end of the night.
In medium resolution,
the lamp calibration was refined each time on the OH sky lines observed 
in the spectra. In addition, the object and standard spectra were carefully   
aligned each other against the atmospheric absorption features before being
rationed. This procedure leads to a
wavelength calibration error in medium resolution mode that we estimated
of the order of  0.1 \AA (i.e. about 1-2 km\,s$^{-1}$).\\

The spectra were obtained during not photometric nights, with a seeing
variable between 1.5 and 2 arcsec. To flux calibrate our spectra we therefore 
used the broad band photometry obtained at SOFI with the NTT instrument
during March 2003, whose values are reported in Table 1. 
For the IRS5 and IRS6
sources, since the NTT resolution is not high enough to resolve the
binaries, we derived the magnitudes of the relative components
using \HH\, and \KK\, VLT-ISAAC images taken from the ESO archive 
(ID programme 63.I-0691, April 1999).
These are also reported in Table 1. From this table, and 
from the comparison of our derived photometry with the photometry given by 
Taylor \& Storey (1984), we notice that a certain degree of variability is present in 
the considered sources, with the largest variation exhibited by HH100 IR and IRS3.
HH100 IR source was already known to be highly variable, with near 
IR magnitudes changing by values as large as more than two magnitudes on 
timescales of one year (Molinari et al. 1993). 

\begin{figure}[!ht]
\resizebox{\hsize}{!}{\includegraphics{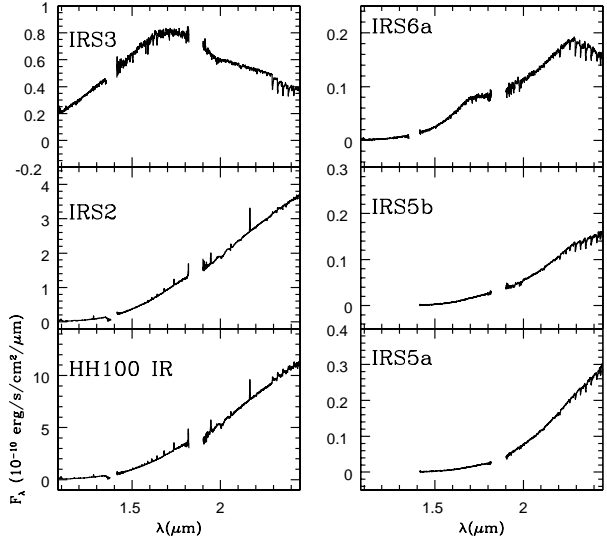}}
   \caption{\label{fig:SED}Low resolution \JJ,\HH\, and \KK\, spectra of 
   the observed sources}
\end{figure}

\begin{figure}
\resizebox{\hsize}{!}{\includegraphics{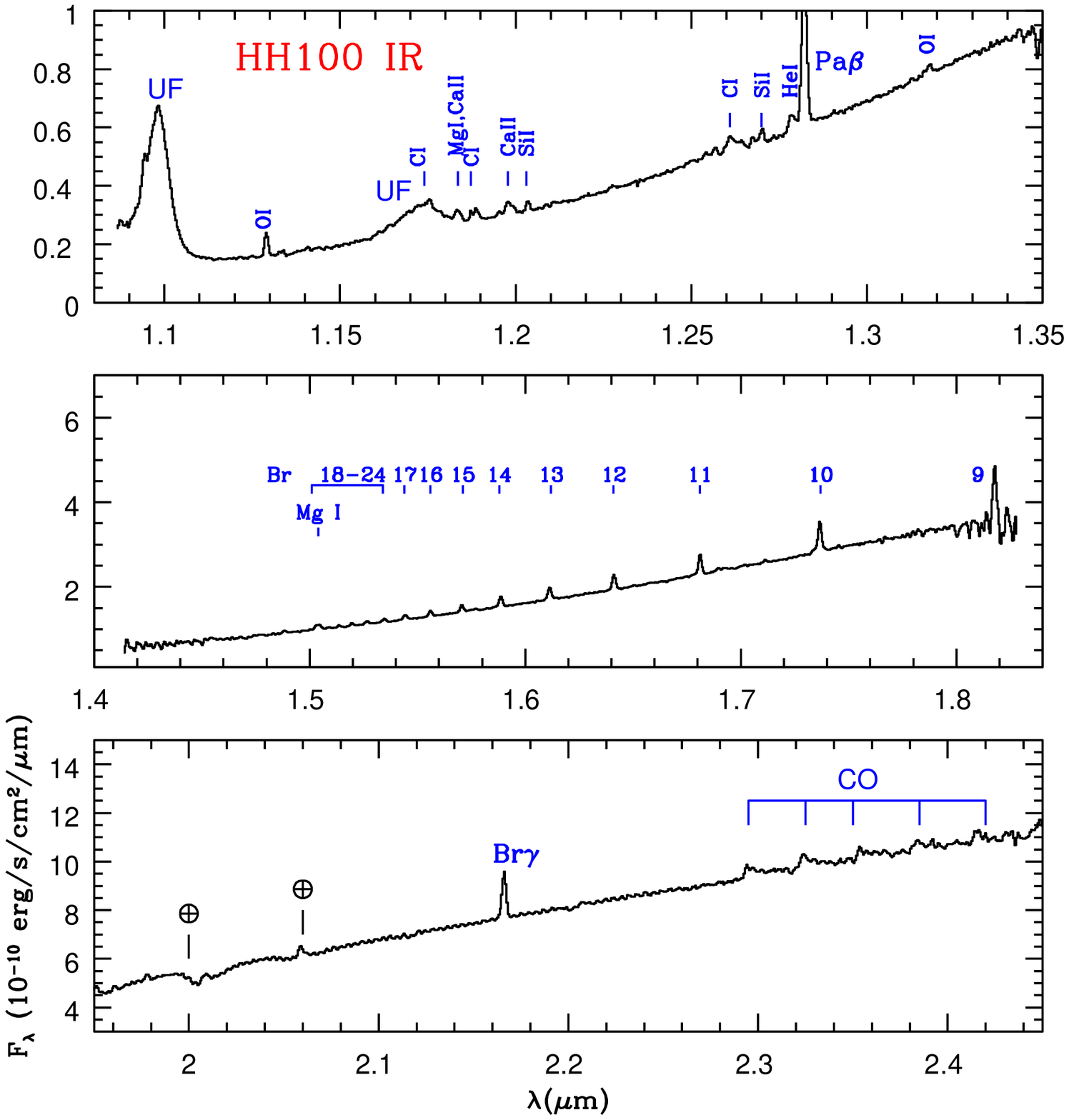}}
   \caption{ \label{fig:HH100LR}Low resolution \JJ,\HH\, and \KK\, spectra of HH100 IR with the main emission lines and features labelled.}
\end{figure}

\begin{figure}
\resizebox{\hsize}{!}{\includegraphics{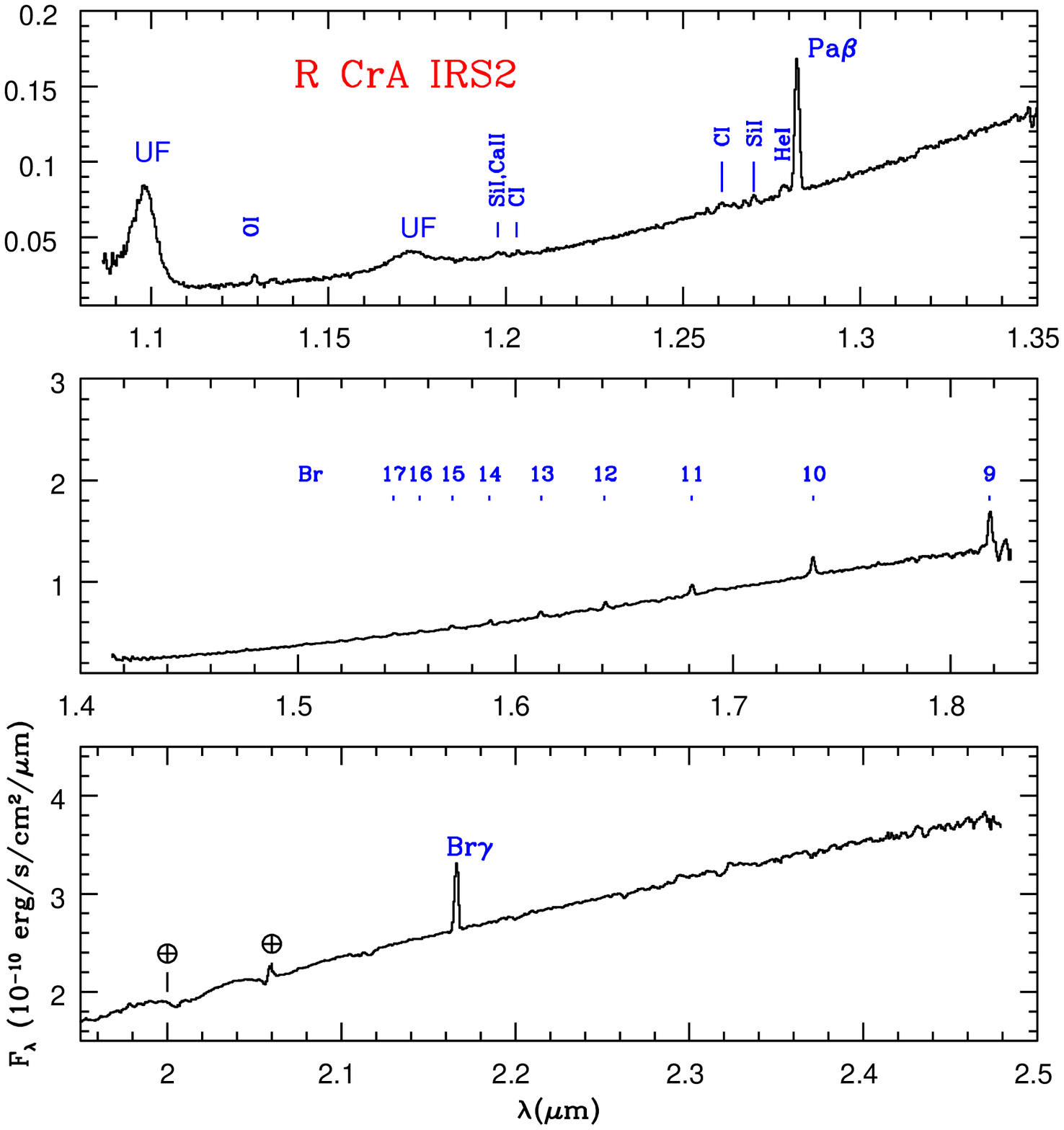}}
   \caption{ \label{fig:IRS2LR}Low resolution \JJ,\HH\, and \KK\, spectra of IRS2 with the main emission lines and features labelled.}
\end{figure}

The complete low resolution spectra of the targetted sources are shown in
Fig. \ref{fig:SED}. The NIR SEDs of all the sources 
but IRS3 are characterized by a continuum steeply rising from the 
\JJ\, to the \KK\, band. The IRS 3 SED peaks in the \HH\, band and then
declines. This evidence,
and the fact that the \HH-\KK\, color 
of this source more recently measured,
is smaller by about 1 mag with respect to the value given by Taylor \& Storey (1984)
(see Table 1),
indicate that the IRS3 IR excess is significantly reduced with time.
In the low resolution spectra of IRS3, IRS5a,b and IRS6a some features in absorption
are present, while no significant line in emission is detected. Conversely,
IRS2 and HH100 IR are characterized by rich spectra in emission 
with no absorption features (see also Figs. \ref{fig:HH100LR},\ref{fig:IRS2LR}).
Moreover in both of these spectra, and to a lesser extent also in IRS6a,
we detect two broad features at $\sim$1.10 and 1.17\um\, which remain
unidentified.

\begin{figure*}
\resizebox{\hsize}{!}{\includegraphics{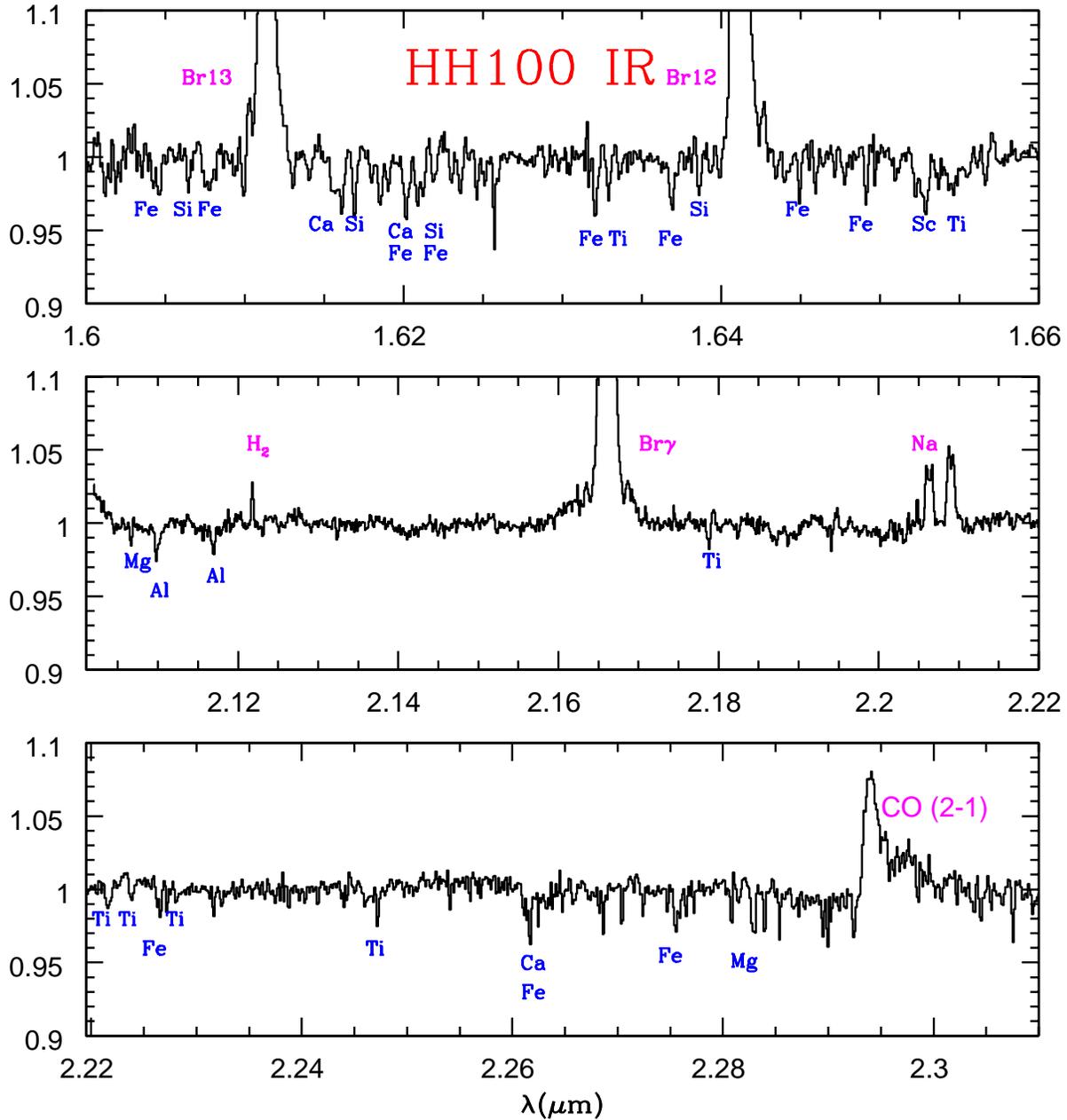}}
   \caption{\label{fig:HH100MR}Continuum normalized, medium resolution spectra of HH100 IR.
   The most important emission and absorption features are labelled.}
\end{figure*}

\begin{figure*}
\resizebox{\hsize}{!}{\includegraphics{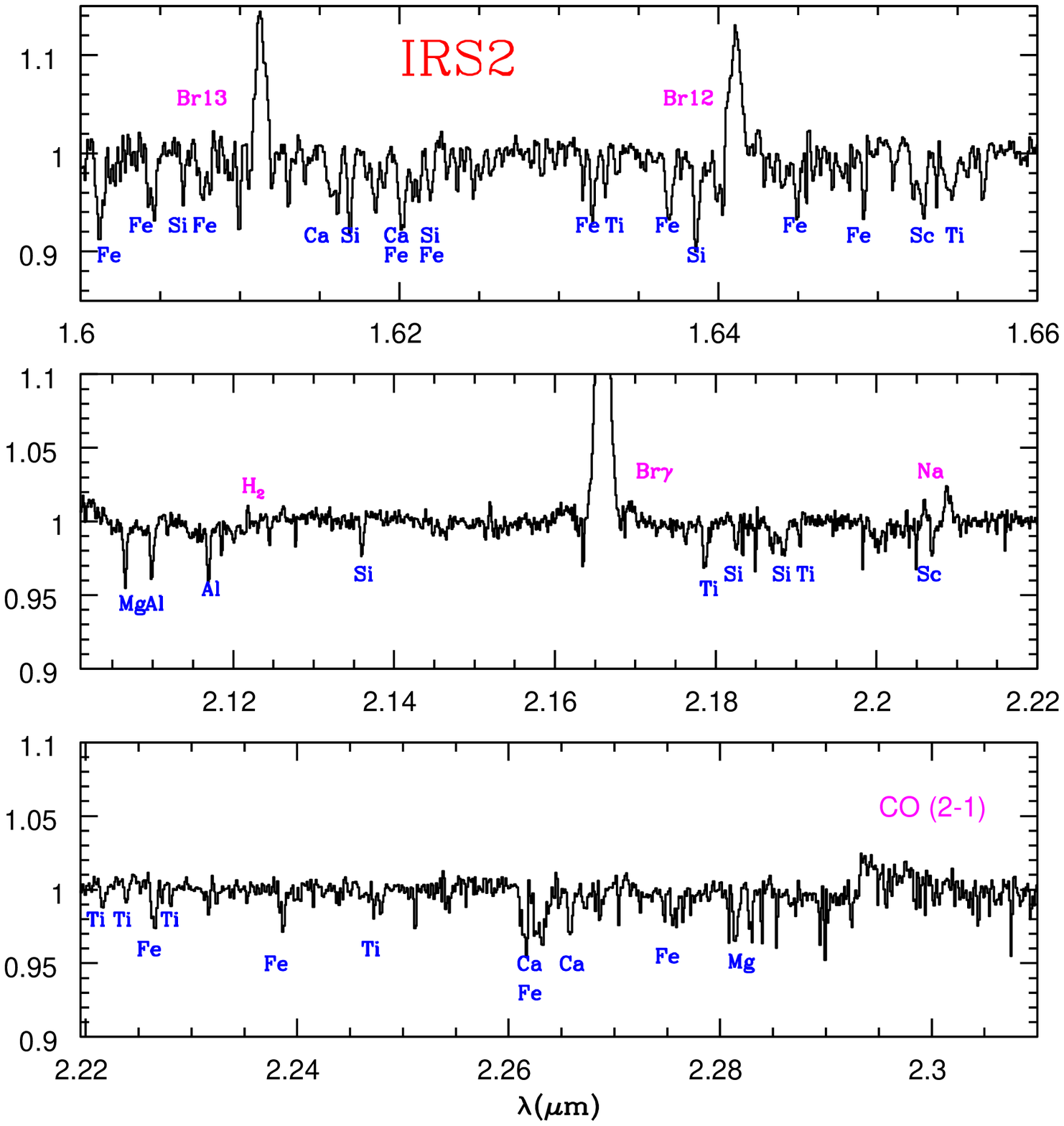}%
\includegraphics{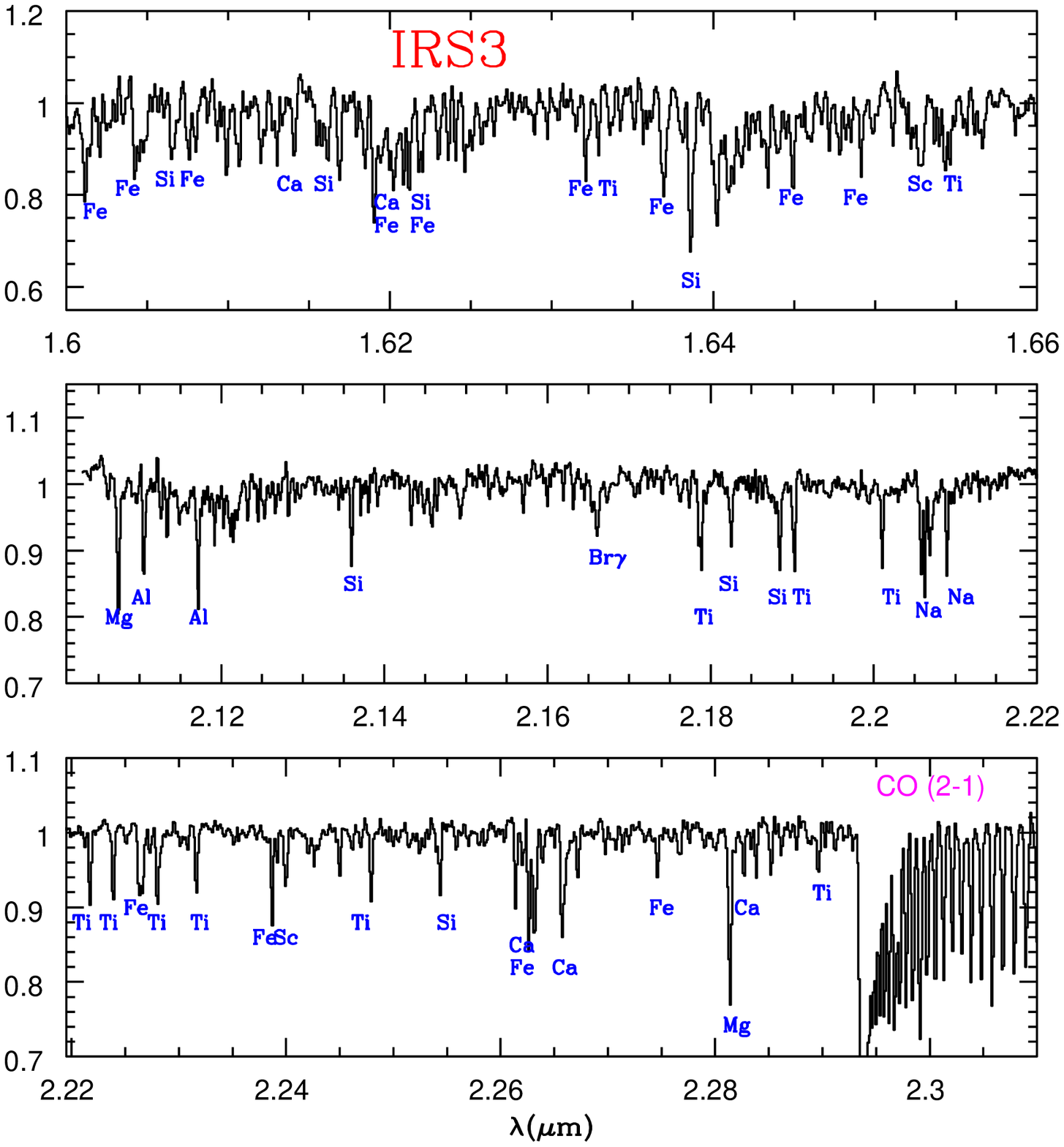}
}
   \caption{\label{fig:IRS2_3MR} As in Fig. \ref{fig:HH100MR} but for IRS2
   (left) and IRS3 (right).}
\end{figure*}

\begin{figure*}
\resizebox{\hsize}{!}{\includegraphics{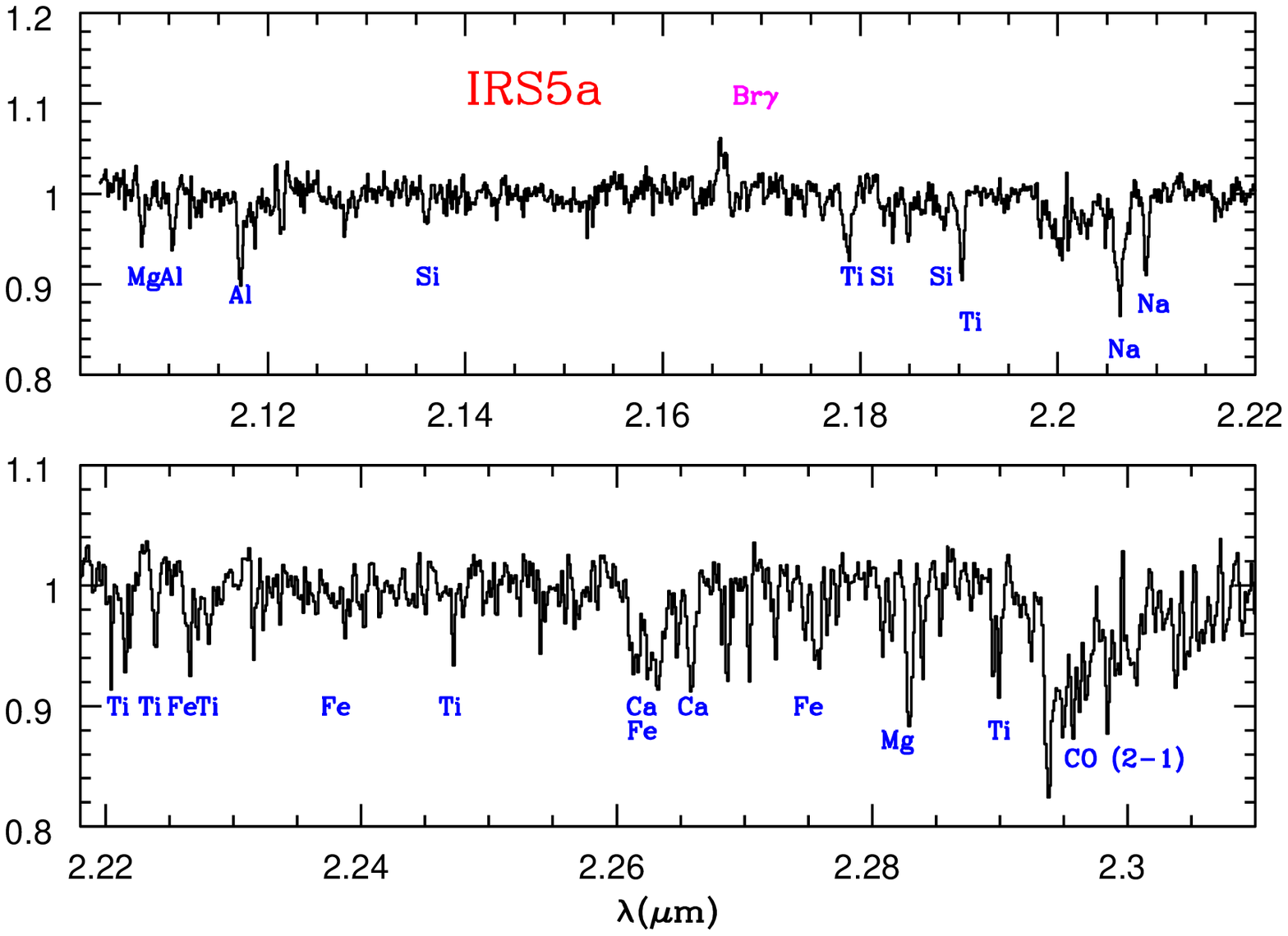}%
\includegraphics{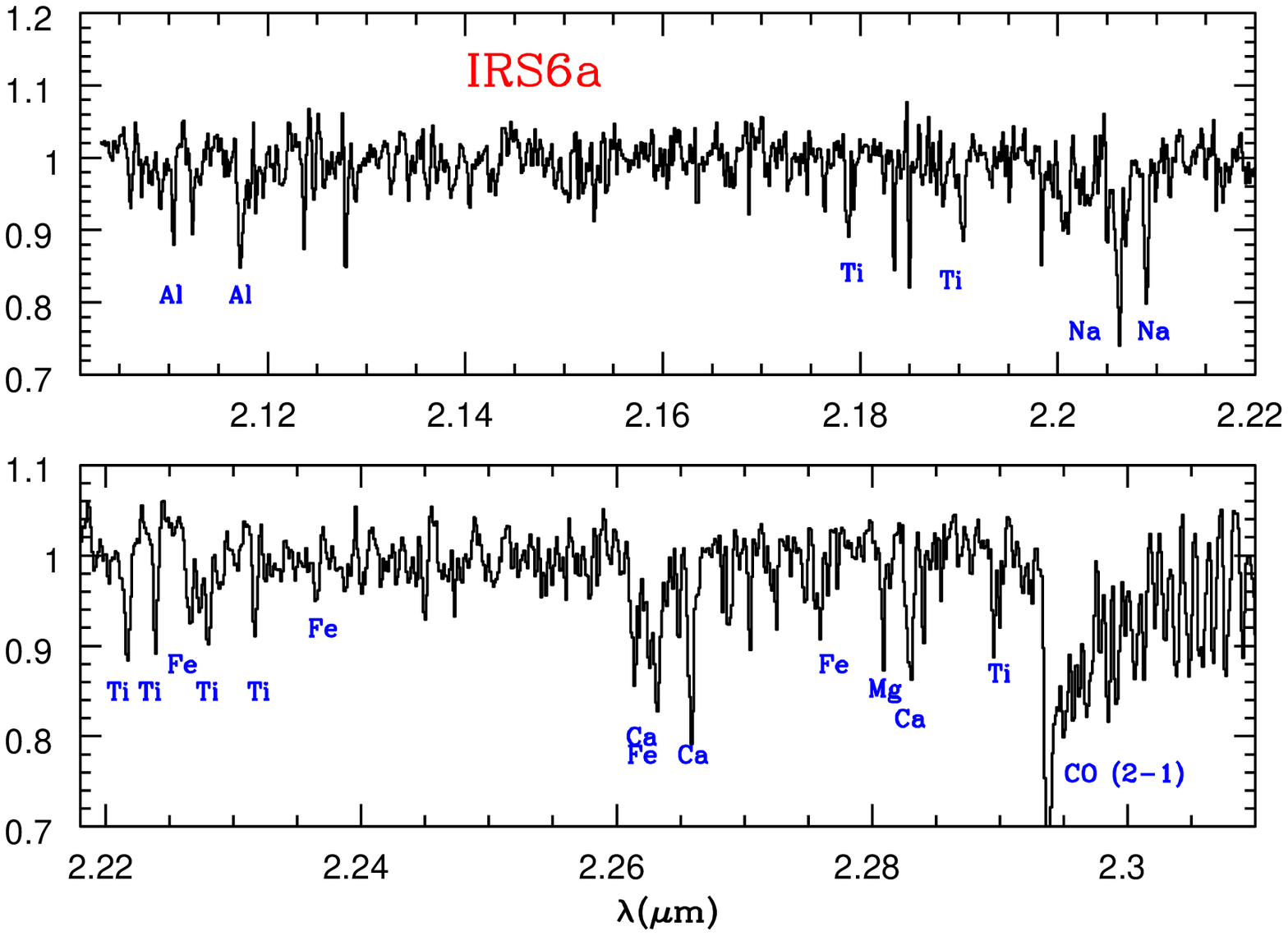}}
   \caption{\label{fig:IRS5_6MR} As in Fig. \ref{fig:HH100MR} but for IRS5a
   (left) and IRS6a (right).}
\end{figure*}

Figs. \ref{fig:HH100MR},\ref{fig:IRS2_3MR} and \ref{fig:IRS5_6MR}
 show the normalized, high resolution spectral 
segments observed in each source. 
Given the weakness of IRS5b with respect to the other component, it was
not possible to extract its medium resolution spectrum 
and therefore only the MR spectrum of IRS5a has been analyzed.
All these spectra are characterized by the
presence of many features in absorption. This is true also for HH100 IR and IRS2
where the higher resolution has allowed to detect narrow and weak features
which remained undetected in the low resolution spectra.
 We will separately discuss the emission and absorption spectra of the
observed sources in the following \S 4 and 5.

\section{The emission lines}

HH100 IR and IRS2 are the only sources showing a prominent emission 
spectrum rich of features (Figs. 3 and 4). 
Such features mainly include permitted transitions of neutral atoms. 
The HI lines  are the strongest of them: we detect Br$\gamma$, Pa$\beta$ 
and other lines from the Brackett series up to n=20 in HH100 IR and 
IRS2, while a weak Br$\gamma$ emission is also mesured in the MR spectrum 
of IRS5a. The HI lines observed in HH100 IR have been already presented and 
discussed in Nisini et al. (2004); in Table 2 we report the fluxes 
and 
the kinematical properties of the HI lines detected both in the low and
high resolution spectra of IRS2 and IRS5a.
A full account on the interpretation of the HI line emission in HH100 
IR has been given in Nisini et al. (2004). 
The Brackett decrement observed on HH100 IR is consistent with emission 
from moderately optically thick lines
originated from a compact region of few stellar radii. 
Moreover, the
fact that the Brackett line profiles observed in high resolution are almost 
symmetrical with linewidths decreasing with the upper quantum number, 
is an indication in favour of an origin in an expanding stellar 
wind rather than in an accretion region, as often suggested for these 
lines (e.g. Muzerolle et al. 1998). 
In such an hypothesis, a mass loss rate of $\sim$10$^{-6}$ M$_{\odot}$\,yr$^{-1}$ 
has been derived for HH100 IR by using a simple model of an ionized wind. 
The Brackett lines observed in 
IRS2 shows characteristics similar to the HI lines of HH100 IR, such 
as the symmetry of the Br$\gamma$ line and large linewidths. The Br12 and 13 
profiles are also narrower than the Br$\gamma$
line, but they appear less symmetric partially due to a superposition 
of few photospheric absorption lines which make difficult to reconstruct the 
intrinsic line profile.
The IRS2 Brackett decrement resulting after having dereddened the observed lines 
with an extinction value $A_{\rm V}$=20 (see Table 7) is compatible with
line optical depths smaller than in the HH100 IR case (see e.g. Fig. 2
of Nisini et al. (2004)), indicating  
lower densities of ionized material.  If interpreted in the same wind model 
adopted for HH100 IR, such a decrement is consistent with a mass loss 
rate of ionized gas of $\sim$10$^{-8}$ M$_{\odot}$\,yr$^{-1}$, 
corresponding to a total mass loss rate 
of $\sim$ 1-2\, 10$^{-7}$ M$_{\odot}$\,yr$^{-1}$ assuming a ionization fraction 
between 0.05 and 0.1.

The other emission lines detected in the low resolution spectra of HH100 IR
and IRS2 are listed 
in Table 3
where we have included only the features having an S/N larger than 3 and linewidth 
comparable to the instrumental resolution.  
In this table we indicate, when possible, the 
most probable identification done using the NIST database\footnote{
http://physics.nist.gov/PhysRefData/ASD1/} and guided by
the spectra of other YSOs (Kelly et al. 1994) or objects with
likely similar excitation conditions (Walmsely et al. 2000, Hamann et al. 
1994). We tried as much as possible to check for the plausibility of 
a line identification by verifying that all the transitions of the 
same atom expected to originate from the same term were also 
detected.
The observed atomic lines include several CI, OI, SiI and Mg I transitions,
all coming from levels with energies $<$ 12.3 eV. The only
ionic lines detected are from singly ionized Fe and Ca, which have 
ionization potential lower than 8 eV.
In the high resolution spectral segments, in addition to the 
Br$\gamma$, Br12 and Br13 lines, the Na I doublet lines at 2.206 and 
2.209$\mu$m, and the 2.12$\mu$m H$_{2}$ v=1-0 S(1) lines are detected 
in emission in HH100 IR and IRS2 and their fluxes and kinematical 
characteristics reported in Tab 3.

Most of the observed atomic lines can be excited by both recombination and 
photo-excitation. For the two OI lines at 1.1287,1.3165 \um\, fluorescence
excitation by either an UV continuum or resonant absorption of
HI Ly$\beta$ photons have been proposed as the most likely excitation
mechanisms (see e.g. Kelly et al. 1994, Walmsley 2000). The several CI lines
are on the other hand the brightest permitted lines predicted by recombination
theory (Escalante \& Victor 1990).  The several lines from SiI and MgI, which have 
low ionization potentials, may also originate from recombination
cascade. 
Some of the atomic species observed are those responsible for strong
permitted optical lines in T Tauri stars, such as NaI and CaII. 
The levels from which the IR lines originate lie at excitation energies
larger than the levels of the optical lines, but it is conciveable that
the origin of the emission must be similar. Among different possibilities,
emission from the disk-star accretion region (Basri \& Bertout 1989, Hartmann 1998), 
from a stellar wind (Hartmann et al. 1990) and from an active cromosphere 
(Calvet et al. 1984) are the most commonly proposed.
The kinematical characteristics of the two Na I lines detected in high 
resolution are remarkably similar to those of the Br$\gamma$ 
emission, both in terms of linewidths and V$_{LSR}$, which suggests
that they come from the same emission region.
We also note the absence of forbidden lines, e.g. from [FeII], which 
indicate that the emission line region is characterized by high densities 
where 
forbidden lines with critical densities lower than 10$^{5}$ cm$^{-3}$
are quenched.

The weak H$_2$ 2.12$\mu$m line detected in HH100 IR and IRS2 (Table 4) may be
an indicator of both shocked gas in the interaction of the stellar wind with 
the ambient medium, and emission from the inner warm region of a circumstellar disk.
The line is blueshifted and only barely resolved, 
with an intrinsic width of $\la$ 20 \kms. Given these characteristics, the detected
H$_2$ emission may trace the base of a compact molecular outflow, such as
those already detected in other Class I sources by Davis et al. (2001).
In our case the emission is not spatially resolved, however the adopted
slit  may not be aligned with the compact jet emission. 
The origin in a circumstellar disk is however equally plausible.
The upper limit on the line widths, together with the source mass estimates 
given in Tab 7,
indicate that for a keplerian disk, the observed line are
consistent with an emission region at $>$ 1 AU from the stellar source. 
Ro-vibrational H$_2$ emission with linewidths similar to those observed by us
have been detected from disks of some T Tauri stars (Bary et al. 2003).
In conclusion both the jet and the disk hypothesis seem to be  plausible for the 
2.12$\mu$m line origin.
Finally CO in emission from different vibrational band heads in the 2.3$\mu$m
region is also detected in HH100 IR and IRS2.  
CO band head emission is a tracer for dense molecular gas at T$\sim$2000-3000 K,
and it is usually interpreted as a sign for the 
presence of active accretion disks. The reason is that to have this band in emission
large columns of warm gas need to be present, and accretion may represent
a viable disk heating mechanism (Najita et al. 1996). 

The spectra of HH100 IR, IRS2 and IRS6a show, in addition to the 
emission lines, also two broad and strong unidentified features in emission, at 1.098 and 
1.173 \um, whose parameters are given in Table 5.
 Such features seem not originated by spurious or
instrumental effects since they are not observed neither in the IRS3 
spectrum nor in the spectra of the standard stars. They do not appear 
spatially resolved along the slit, being confined on the source. 
The large widths and strengths of these features resamble the characteristic
shape of PAH emission, which however are so far never been observed at 
these wavelengths. The observations of PAH at such short wavelengths 
would imply electronic excitation of these macro-molecules and 
consequently the presence of strong UV fields which seem not to be 
associated with the low luminosity sources we are considering. The attribution
of these features to PAH emission is also not favoured by the non-detection
of the vibrationally excited PAH feature at 3.3\um\, in the spectrum
of HH100 IR (Whittet et al. 1996).
Diffuse interstellar bands (DIBs) have been detected in the considered
spectral range (Joblin et al. 1990) but they are features usually observed
in absorption against a strong continuum. Unidentified dust emission
features at 1.15 and 1.5\um\, have been finally observed towards the reflection
nebula NGC7023 (Gordon et al. 2000). Such features are however much broader
($\sim$ 0.2-0.3\um\,) and more diffuse than those detected by us.


\begin{table*}
\label{tab:em_lines}
\caption[]{Emission lines observed in HH100 IR and IRS2 (excluding HI
lines)}
\vspace{0.5cm}
    \begin{tabular}[h]{cccl}
      \hline \\[-5pt]
$\lambda_{obs}$ &  $F({\Delta F})$ & $F({\Delta F})$ &
Identification ($\lambda_{vac}$ in $\mu$m)\\
 $\mu m$  & \multicolumn{2}{c}{10$^{-15}$ erg\,s$^{-1}$\,cm$^{-2}$ } &\\
\hline\\[-5pt]
  &    HH100 IR & IRS 2\\
\hline\\[-5pt]
 1.1291 &  1.52(0.05) & 0.89(0.07)& OI $3p^3P$-$3d^3D^0$ 1.1287\\
 1.1327 &  0.05(0.05) & ...       & CI $3d^1D^0$-$3p^1P$ 1.1333\\
 1.1339 &  0.15(0.05) & ...       & CI $3d^1F^0$-$3p^3D$ 1.1339\\
 1.1641 &  0.34(0.08) & ...       & CI $3d^3D^0$-$3p^3D$ 1.1632, 1.1651\\
 1.1754 &  0.59(0.08) & ...       & CI $3d^3F^0$-$3p^3D$1.1751,1.1756, 1.1758\\
 1.1831 &  0.37(0.08) & ...       & MgI $4s^1S$-$4p^1P^0$ 1.1831\\
 1.1839 &  0.55(0.08) & ...       & ?\\
 1.1835 &  0.84(0.08) & ...       & CaII $5p^2P^0$-$5s^2S$ 1.1842\\
 1.1873 &  0.27(0.08) & ...       & ?\\
 1.1888 &  0.84(0.08) & 0.27(0.06)& CI $4s^3P^0$-$3p^3D$1.1883\\
 1.1954 &  0.32(0.05) & ...       & CaII $5p^2P^0$-$5s^2S$ 1.1953 \\
 1.1978 &  1.02(0.05) & 0.83(0.06)& ?\\
 1.1992 &  0.81(0.05) & 0.20(0.06)& SiI $4p{^3D}$-$4s^3P^0$ 1.9948\\
 1.2035 &  0.60(0.05) & 0.35(0.06)& SiI $4p^3D$-$4s^3P^0$ 1.2035\\
 1.2086 &  0.19(0.05) & ...       & MgI $4f^1F^0$-$3d^1D$ 1.2087\\
 1.2111 &  0.62(0.05) & ...       & SiI $4p^3D$-$4s^3P^0$ 1.2107\\
 1.2275 &  0.30(0.04) & ...       & SiI $4p^3D$-$4s^3P^0$ 1.2274\\
 1.2542 &  0.26(0.04) &...        & \\
 1.2568 &  0.36(0.04) &0.3(0.1)   & CI $3d^3P^0$-$3p^3P$1.2566 + \\
        &             &           & [FeII] $a^6D_{9/2}$-$a^4D_{7/2}$ 1.2567\\
 1.2611 &  1.93(0.04) & 1.6(0.1)  & CI $3d^3P^0$-$3p^3P$1.2617\\
 1.2640 &  0.93(0.04) & 0.8(0.1)  & FeII $z^6F_{7/2}$-$c^4F_{5/2}$ 1.2646\\
 1.2675 &  0.60(0.04) & 0.5(0.1)  & ?\\
 1.2701 &  1.02(0.04) & 1.0(0.1)  & ?\\
 1.2784 &  1.18(0.04)& 1.2(0.2)   & ?\\
 1.3176 &  0.64(0.12)& ...        & OI $3p^3P$-$4s^3S^0$ 1.3166\\
 1.4884 &  10.3(1.3) & ...        & MgI $4f^3F^0$-$3d^3D$ 1.4882\\
 1.5047 & 20.1 (0.7) & 3.5 (0.9)  & MgI $4p^3P^0$-$4s^3S$ 1.5044 + Br23\\
 1.7113 &  11.1(1.8) & ...        & MgI $4p^1P^0$-$4s^1S$ 1.7113\\
 2.294  & 540(30) & 87(17)        & CO v=2--0\\
 2.324  & 600(30) & 110(25)       & CO v=3--1\\
 2.354 & 500(30) & ...            & CO v=4--2\\
 2.324  & 390(40) & ...           & CO v=5--3\\
 2.324  & 360(40) & ...           & CO v=6--4\\
\hline \\[-5pt]
\hline \\[+5pt]
      \end{tabular}
\end{table*}


\begin{table}
\label{tab:HI_lines}
\caption[]{HI lines in IRS2 and IRS5a}
\vspace{0.5cm}
\begin{tabular}{cccc}
\hline\\[-5pt]
\multicolumn{4}{c}{Fluxes}\\
\hline\\
 & & IRS 2 & IRS5a\\
  $\lambda_{obs}$& Line& $F({\Delta F})$ &  $F({\Delta F})$ \\
$\mu$m  & & \multicolumn{2}{c}{10$^{-15}$ erg\,s$^{-1}$\,cm$^{-2}$ } \\[+5pt]
\hline\\
1.282 & Pa$\beta$ & 13.7(0.4) &\\
1.519 & Br 20 & 1.6(0.5) &\\
1.527 & Br 19 &  2.9(0.7) &\\
1.534 & Br 18 &  3.7(1.1) &\\
1.544 & Br 17 &  4.2(0.2) &\\
1.556 & Br 16 & 4.1(0.4) &\\
1.571 & Br 15 & 7.2(0.1) &\\
1.588 & Br 14 & 6.0(0.7) &\\
1.612 & Br 13 & 10.7(0.1) &\\
1.641 & Br 12 &  9.3(0.6) &\\
1.681 & Br 11 & 19.1(0.1) &\\
1.737 & Br 10 & 41.5(0.9) &\\
1.818 & Br 9 &  82.3(1.7) &\\
2.166 & Br$\gamma$ & 181.0(5.4) & 1.3(0.1)\\
\hline\\[-5pt]
\end{tabular}
\begin{tabular}{cccccc}
\hline\\[-5pt]
\multicolumn{6}{c}{Kinematical information}\\
\hline\\[-5pt]
 & & \multicolumn{2}{c}{IRS 2} & \multicolumn{2}{c}{IRS5a}\\
  $\lambda$ & Line& $V_{LSR}$ & $\Delta~V$ & $V_{LSR}$ & $\Delta~V$\\
  ($\mu$m) & & km\,s$^{-1}$ & km\,s$^{-1}$ & km\,s$^{-1}$ & km\,s$^{-1}$\\[+5pt]
\hline\\[-5pt]
1.6113 & Br 13      & -13$\pm$4 & 130 & &\\
1.6411 & Br 12      & -11$\pm$4 & 146 & &\\
2.1661 & Br$\gamma$ & -9$\pm$4 & 180 & 0$\pm$10 & 122\\
\hline\\[-5pt]
\end{tabular}

\end{table}

\begin{table*}
\label{tab:em_MR}
\caption[]{Other emission lines in the high resolution spectra}
\vspace{0.5cm}
\begin{tabular}{cccccccc}
\hline\\[-5pt]
 & & \multicolumn{3}{c}{HH100 IR} & \multicolumn{3}{c}{IRS2}\\
  $\lambda$ & Line& $V_{LSR}$ & $\Delta~V$ & $F\pm\Delta~F$ & $V_{LSR}$ & $\Delta~V$ & $F\pm\Delta~F$\\
  $\mu$m & & km\,s$^{-1}$ & km\,s$^{-1}$ & 10$^{-17}$ erg\,s$^{-1}$\,cm$^{-2}$
  &km\,s$^{-1}$ & km\,s$^{-1}$&  10$^{-17}$ erg\,s$^{-1}$\,cm$^{-2}$ \\[+5pt]
\hline\\[-5pt]
2.1218 & H$_2$      & -6  & $\la$20  & 5.4 $\pm$ 0.3& -6 & $\la$20 &3.4 $\pm$ 1.0\\
2.2062 & Na I $^2P^{0}_{3/2}$--$^2S_{1/2}$ &5 &142 & 32.0 $\pm$ 0.8& ... & ... &8.8 $\pm$ 1.5\\
2.2090 & Na I $^2P^{0}_{1/2}$--$^2S_{1/2}$ &2 &133 & 42.0 $\pm$ 1.5& -11 & 117 &19.0 $\pm$ 0.5\\
\hline\\[-5pt]
\end{tabular}
\end{table*}

\begin{table*}
\label{tab:UIF}
\caption[]{Parameters of the 1.1$\mu m$ unidentified features}
\vspace{0.5cm}
    \begin{tabular}[h]{ccccc}
      \hline \\
          & \multicolumn{2}{c}{HH100 IR} & \multicolumn{2}{c}{IRS2}\\
$\lambda$ & $F \pm {\Delta F}$ & FWHM & $F \pm {\Delta F}$ & FWHM  \\
 $\mu m$ &  10$^{-14}$ erg\,s$^{-1}$\,cm$^{-2}$ & $\mu m$ &  10$^{-14}$ erg\,s$^{-1}$\,cm$^{-2}$ & $\mu m$ \\
\hline\\
 1.098 & 42.9$\pm$0.9 & 7.4\,10$^{-3}$ &5.1$\pm$0.2 & 7.5\,10$^{-3}$ \\
 1.173 & 13.2$\pm$0.7 & 1.3\,10$^{-2}$ &1.4$\pm$0.1 & 1.2\,10$^{-2}$ \\
\hline \\
      \end{tabular}
\end{table*}

\section{The absorption features}
In the two \KK\, band spectral segments, the main absorption features observed
 are those of lines from neutral atoms like Na, Al, Fe, Mg, Ti and Si,
as well as the CO 2-0 band; these are the 
typical features expected from the photosphere of late type (K-M) stars 
(e.g. Wallace \& Hinkle 1996, 1997) and commonly observed in pre-main sequence
sources (e.g. Luhman \& Rieke 1998). In HH100 IR and IRS2 such features 
are detected at only 1-5\% of the continuum level. 
In the \HH\, band spectral segment, several more features are detected, 
most of which are very weak. The most significant of these features are
indicated in Figs. \ref{fig:HH100MR},\ref{fig:IRS2_3MR} 
and \ref{fig:IRS5_6MR} and have been identified through the spectra of standard
stars of
Meyer et al. (1998) and the Kurucz atomic spectral line database \footnote{
1995 Atomic Line Data (R.L. Kurucz and B. Bell) Kurucz CD-ROM No. 23. 
Cambridge, Mass.: Smithsonian Astrophysical Observatory}.
The absorption lines are resolved, with widths of several tens \kms\,
which vary depending on the considered feature in the same objects. 
Such a broadening is due to different contributions, including rotation
and Zeeman broadening, since these young stars are expected 
to be fast rotators and to have strong stellar magnetic fields. 
More specifically, the width of lines selected for having small
Land\'e factors and not blended with other features are of the order of $\sim$20\kms\,
in IRS3, 40\kms\,in IRS2, HH100IR and IRS6a, and 60\kms\,in IRS5a. This indicates 
that all the sources but IRS3 possess large $v\,sin\,i$ values, comparable 
with the projected rotational velocities estimated in other flat-spectrum and Class I
sources and higher than the average $v\,sin\,i$ values measured in T Tauri stars
($<v\,sin\,i>$ $\sim$15\kms, Bouvier et al. 1986).

\subsection{Spectral type and veiling measurement}

In the assumption that the observed absorption lines originate from stellar
photosphere, their equivalent width (EW) should be sensitive to the stellar 
effective temperature and gravity. In the sources we are considering, however,
the line EWs are strongly affected
by the presence of continuum excesses, which can be quantified through
the {\it veiling} parameter, defined as the flux at a given
wavelength in excess with respect to the photospheric flux
($r_{\lambda}=F_{ex,\lambda}/F_{*,\lambda}$).
The veiling is actually a parameter that we also want to derive from the
data, giving us information on the circumstellar activity of the stars. 
The ratios of different lines close in wavelengths is not largely affected
by the veiling and can thus be used for defining the spectral type. Then, 
the absolute value of the line EWs can be in turn compared with the intrinsic EW of spectral
standard lines to estimate the IR veiling. Several diagnostic lines, mainly 
located in the \KK\, band,
have been identified by different authors as particulary suited for spectral classification
(e.g. Luhman \& Rieke 1998).
Among them, different features lying in the region around the Na I doublet at 
2.206$\mu$m are good diagnostic tools for the effective temperature,
while the CO 2-0 bandhead is particularly sensitive to the gravity (e.g. Doppmann \& Jaffe 2003).
However, we have seen that in HH100 IR and IRS2 both the Na I doublet lines and
the CO band-heads are seen in emission instead of absorption, and therefore
cannot be used for spectral classification.
Therefore, to derive
the stellar properties of our sources, we have used different and 
close in wavelength diagnostic lines, mainly located in the \KK\, band.
Details on the considered lines are given in Appendix A.
In practice to derive the spectral type and luminosity class,
we have compared the ratios of selected absorption lines observed in our sources,
with the same ratios as measured in a grid of standard star spectra, selecting 
the standard which minimizes 
the scatter among the different ratios.

\begin{table}
\label{tab:ST}
\caption[]{Spectral types and veiling}
\vspace{0.5cm}
    \begin{tabular}[h]{ccccc}
      \hline \\[-5pt]
Source  & ST  & $T_{eff}$ & $r_{K}$ & $r_{H}$\\[+5pt]
      \hline \\[-5pt]
IRS2 & K2V& 4900$\pm$200&2.9$\pm$0.5&1.4$\pm$0.2\\
IRS5a & K5-K7V&4200$\pm$200&1.0$\pm$0.1 &0.1$\pm$0.1\\
IRS6a &M2V &3580$\pm$100& 0.1$\pm$0.1 & 0.1$\pm$0.1\\
HH100 IR &K5-M0V&4060$\pm$300& 6.0$\pm$0.5&3.5$\pm$0.2\\
IRS3 &K5-M0 III&3800$\pm$100&0.2$\pm$0.1& 0.1$\pm$0.1\\
\hline \\[+5pt]
      \end{tabular}
\end{table}

\begin{table*}
\label{tab:SP}
\caption[]{Stellar parameters}
\vspace{0.5cm}
    \begin{tabular}[h]{ccccccccc}
      \hline \\[-5pt]
Source  & L$_*$  & R$_*$  & M$_*$ & age  & E(H-K) & A$_{V}$ & L$_{acc}$ & $\dot{M}_{acc}$  \\[+5pt]
        & L$_\odot$ & R$_\odot$ & M$_\odot$ & $\times$10$^{6}$ yr & & mag & L$_\odot$ & M$_\odot$\,yr$^{-1}$\\[+5pt]
      \hline \\[-5pt]
IRS2 & 4.3$\pm$1.5 & 2.9$\pm$0.6 & 1.4$\pm$0.3$^a$ & 0.5-1$^a$ & 1.3$\pm$0.2 & 20$\pm$3 & 7.7$\pm$2.5 & 3\,10$^{-7}$$^a$     \\
     &             &             & 1.8$^b$         & 2$^b$     &             &          &             & 2\,10$^{-7}$$^b$    \\
IRS5a & 1.6$\pm$0.5 & 2.4$\pm$0.4 & 0.5$\pm$0.1$^a$ & 0.3-0.5$^a$ & 2.9$\pm$0.2 & 45$\pm$3 & $\sim$0.4 & $\sim$3\,10$^{-8}$$^a$     \\
     &             &             & 0.9$^b$         & 1.3$^b$     &             &          &             & $\sim$2\,10$^{-8}$$^b$   \\
IRS6a & 0.5$\pm$0.2 & 1.5$\pm$0.3 & 0.3$\pm$0.1$^a$ & 0.5-1$^a$ & 1.9$\pm$0.3 & 29$\pm$5 & $<$0.1 & $<5\,10^{-9}$\\
     &             &             & 0.4$^b$         & 1.8$^b$     &             &          &             &    \\
HH100 IR & 3.1$\pm$0.9 & 3.6$\pm$0.7 & 0.4$\pm$0.1$^a$ & 0.1$^a$ & 1.9$\pm$0.2 & 30$\pm$3 & 12$\pm$2 & 2\,10$^{-6}$$^a$     \\
     &             &             & 0.75$^b$        & 0.6$^b$    &             &          &             & 1\,10$^{-6}$$^b$   \\
IRS3 & 0.3$\pm$0.1 & 1.2$\pm$0.1 & 0.5$\pm$0.1$^a$ & 1-5$^a$& 0.2$\pm$0.2 & 10$\pm$3 & $<$0.1 & ... \\
     &             &             & 0.5$^b$        & 4$^b$    &             &          &             &    \\
\hline \\[+5pt]
      \end{tabular}
      
~$^{a}$ Masses and ages derived from the D'Antona \& Mazzitelli (1997) tracks.\\
~$^{b}$ Masses and ages derived from the Siess, Dufour \& Forestini (2000) tracks.
\end{table*}

A difficulty in this procedure is given by
the limited number of available spectra of standard stars at a resolution similar
to that adopted for our observations. For our analysis
 we have mainly used the Wallace \& Hinkle (1997) (for \KK\, band), 
and Meyer et al. (1998) (for the \HH\, band) standard samples,
which are however at resolution 3000, thus not always allowing to compare lines
resolved in our spectra but blended in the standard stars spectra. Moreover,
the Wallace \& Hinkle (1997) sample lacks spectra of dwarfs stars between K5 and 
M2. We therefore also used the dwarfs standard star spectra obtained by 
Greene \& Lada (2002) which have a finer grid between K4V and M2V at a 
resolution  of about 15000, but do not enterely overlap with our observed 
spectral range. 

\begin{figure}
\resizebox{\hsize}{!}{\includegraphics{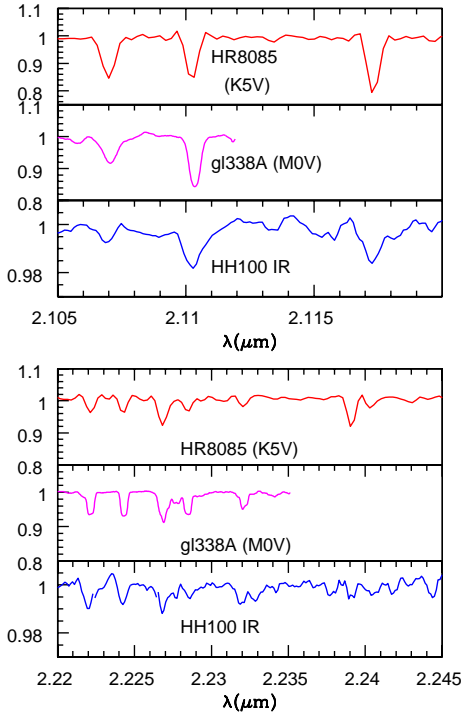}}
   \caption{\label{fig:comp1} Normalized spectral segments of HH100 IR
   in the \KK\, band compared with the spectra of a K5V (Wallace \& Hinkle 1997)
  and a M0V (Greene \& Lada 2001)
   standard stars. }
\end{figure}

\begin{figure*}
\resizebox{\hsize}{!}{\includegraphics{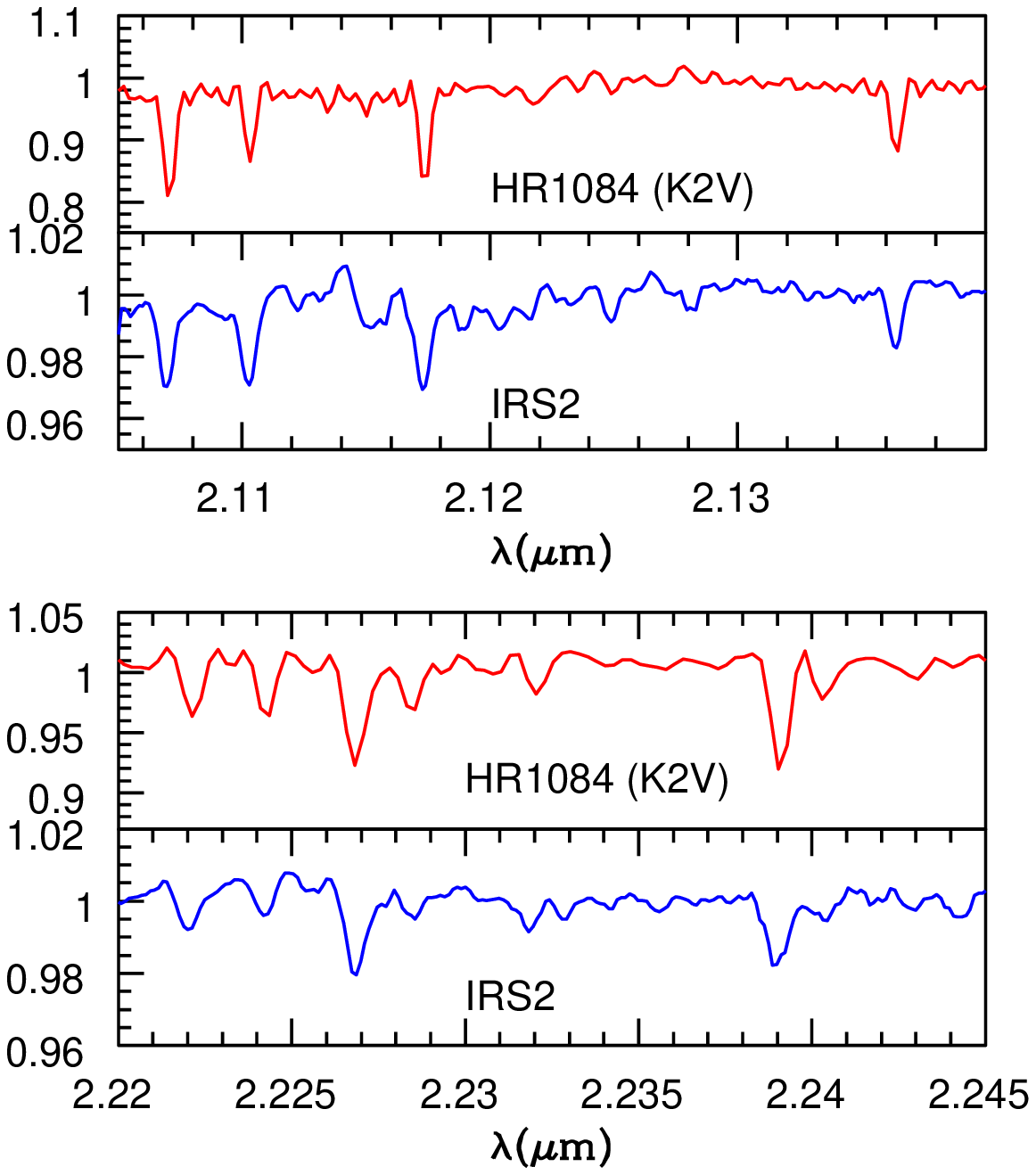}%
\includegraphics{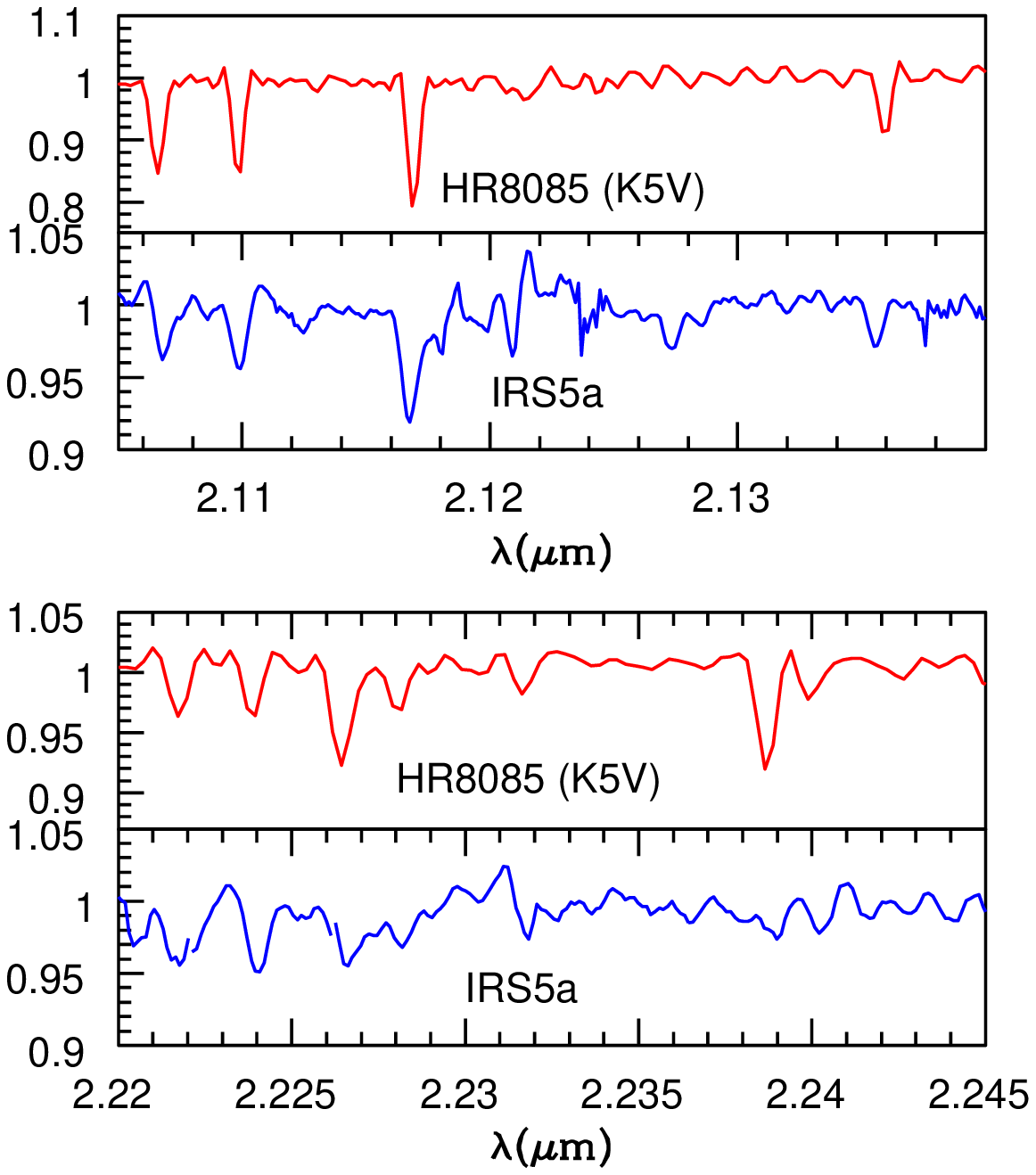}}
\caption{\label{fig:comp2} Normalized spectral segments of IRS2 (left)
   and IRS5 (right)
   in the \KK\, band compared with the spectrum of the standard star
   which better match the observations. The observed spectra have been
   rebinned at the resolution of the standard ($\sim$3000) for comparison.}
\end{figure*}

\begin{figure*}
\resizebox{\hsize}{!}{\includegraphics{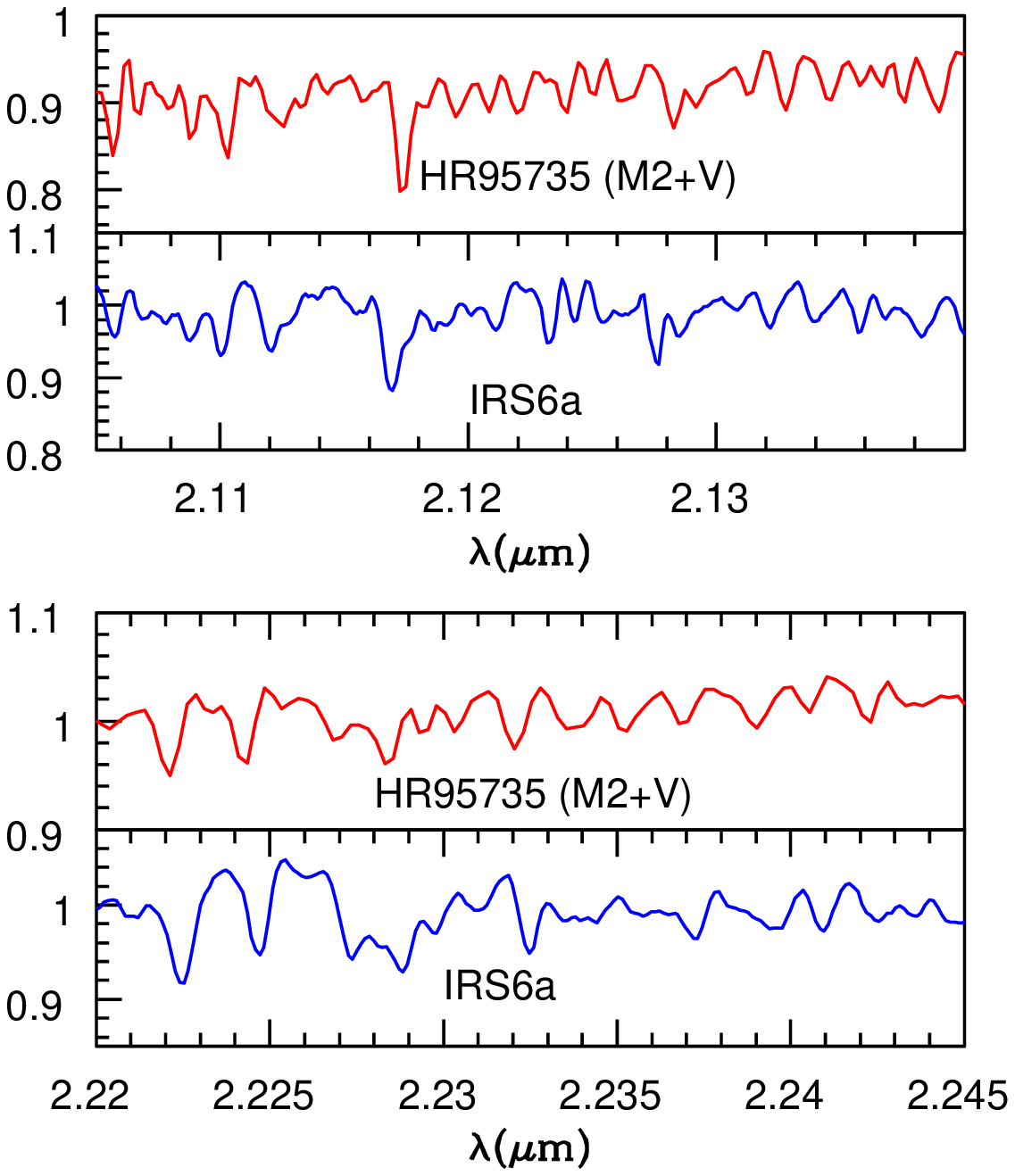}%
\includegraphics{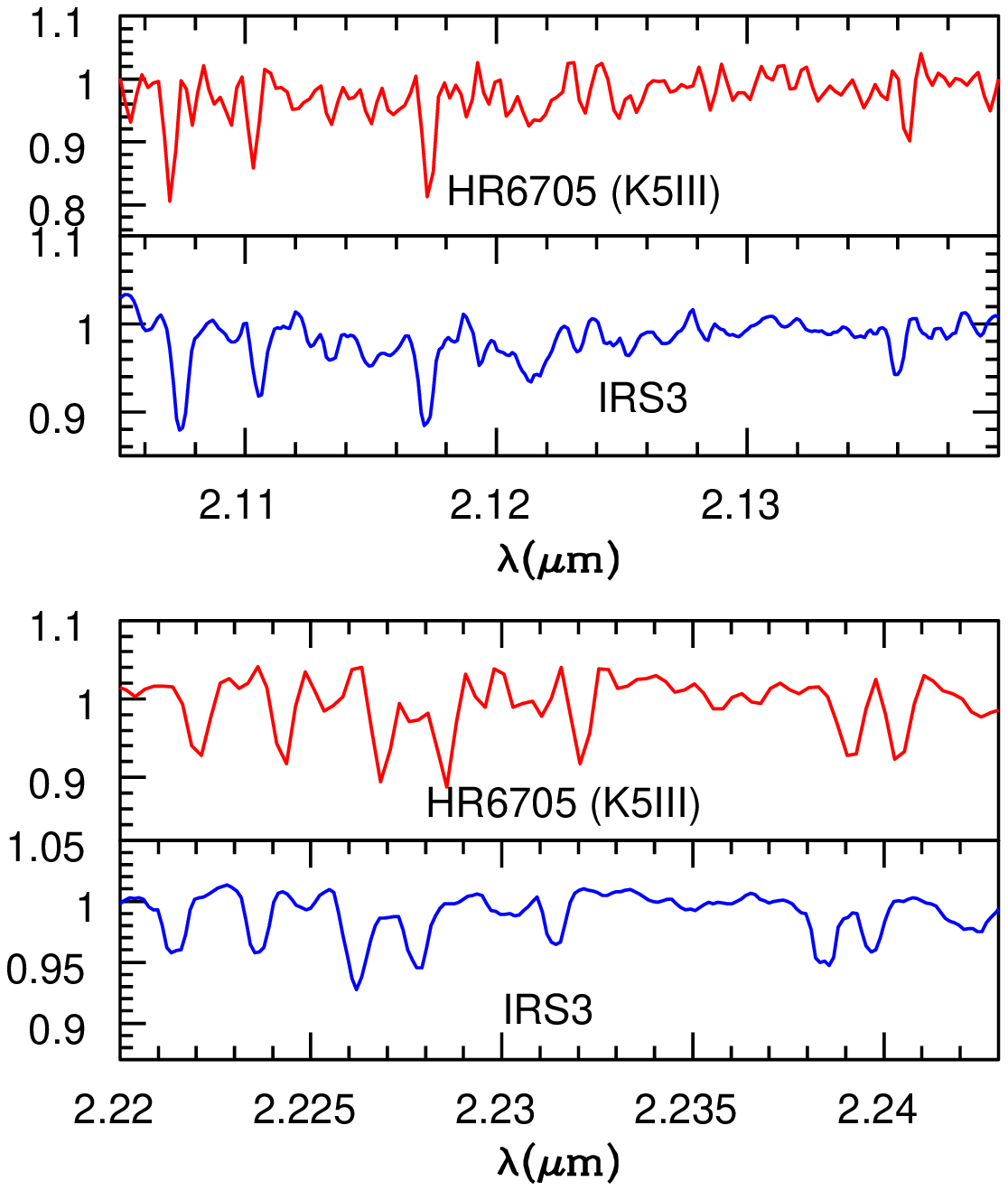}}
\caption{\label{fig:comp3} Normalized spectral segments of IRS6 (left)
   and IRS3 (right)
   in the \KK\, band compared with the spectrum of the standard star
   which better match the observations. The observed spectra have been
   rebinned at the resolution of the standard ($\sim$3000) for comparison.}
\label{plots}  
\end{figure*}

The spectral types and luminosity classes, derived for the sources of our 
sample using the procedure described above, are given in Table 6.
The effective temperatures $T_{eff}$ relative to each 
spectral type have been derived adopting the conversion of spectral type 
into $T_{eff}$ given in Kenyon \& Hartmann (1995).
We estimate that our classification is correct inside one
spectral sub-class, with the exception of HH100 IR where the paucity and weakness
of the absorption features do not allow a classification better than inside 
two spectral sub-classes.
Figs \ref{fig:comp1},\ref{fig:comp2} and \ref{fig:comp3}
show a comparison between the spectra of our sources and the spectra of
standard stars which more closely match the observed spectra.
Most of the sources are consistent with being dwarfs,
as also found for other low mass YSOs (i.e. Greene \& Lada 1997). 
The dwarf spectra shown by young stars have been recognized as an evidence for 
assuming that the observed features originate from the stellar photosphere more
than from a circumstellar disk. Disk photospheres are indeed expected to have
low gravities and thus should resemble giant or supergiants spectra (Greene \& Lada 1997).

Once defined the spectral type, the measured EWs of selected lines, compared with 
the EWs of standard spectra of the corresponding spectral type, have been used to 
derive the veiling in the \KK\, and \HH\, band. For this 
purpose
we used lines not blended at the lower resolution of the available standard spectra, and not
contaminated by possible spurious atmospheric feature residuals. We also selected
lines not too sensitive to the effective temperature to minimize the uncertainty
on the spectral type determination and also not too sensitive to the 
effect of stellar magnetic fields, like the Ti lines at 2.23\um\, whose
EW can be increased by the Zeeman effect due to their high 
Land\'e factor (e.g. Johns-Krull \& Valenti 2000). 
In practice, good lines for the veiling 
determination resulted to be, in addition to the Na line at 2.209$\mu$m, when
observed in absorption, Al at 2.1099\um\, and Ca at 2.2658\um\, for the \KK\, band, 
and Fe (1.6369\um), Si(1.6386\um), in the \HH\, band. For IRS5a and IRS6a 
we do not have the medium resolution \HH\, band spectra, but few 
absorption features are detected in their low resolution spectra. 
Among them, Mg I at 1.7112\um\, is the most suited to evaluate the \HH\,
veiling, being not affected by strong blendings even at low resolution.

The resulted veiling values are also listed in Table 6. From the table,
it can be seen that, as expected, HH100 IR and IRS2 have the largest veiling values, 
$r_{K} \sim$6 and 3 respectively. 
The other considered sources show lower or absent veiling values.
Finally, the veiling in the \HH\, band is about a factor of two
smaller than the $r_{K}$ value. These veiling determinations will be discussed 
in Section 7.3.



\section{Stellar and circumstellar parameters}
\begin{figure}
\resizebox{\hsize}{!}{\includegraphics{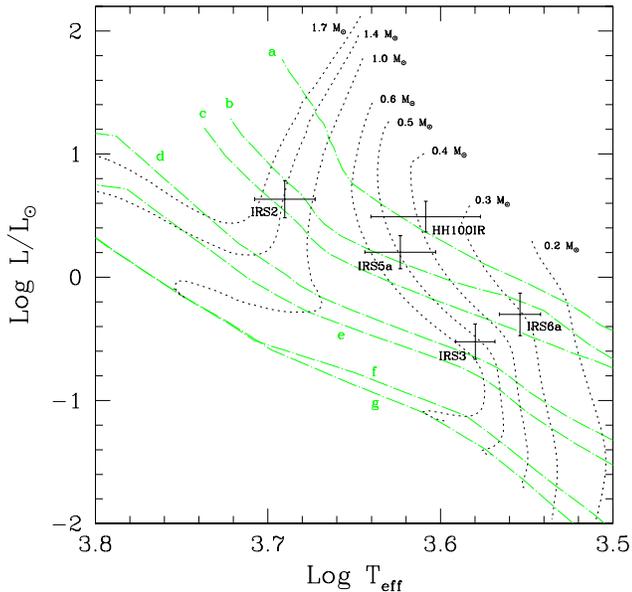}}
   \caption{\label{fig:HR} HR digram of the R CrA sources with $T_{eff}$ and
   stellar luminosity derived from the analysis of the medium resolution IR spectra.
   Evolutionary tracks (short dashed lines) and isochrones (dot-dashed lines)
   from D'Antona \& Mazzitelli (1997) are shown for stellar masses
   between 0.2 and 1.7 M$_{\odot}$. Isochrones are reported for 
   a - 10$^{5}$ yr, b - 5\,10$^{5}$ yr, c - 10$^{6}$ yr, d - 5\,10$^{6}$ yr, e - 10$^{7}$ yr,
   f - 5\,10$^{7}$ yr and g - 10$^{8}$ yr}
\end{figure}

The derivation of spectral type and veiling allows us to estimate other stellar
and circumstellar properties from the available \KK\, and \HH\, photometry
and source bolometric luminosity. 
The infrared color excess $E(H-K)$ can be derived from the observed $(H-K)$ color,
the star color relative to the adopted spectral type $(H-K)_*$ and 
the \rk\, and \rh\, values:

\begin{equation}
E(H-K) = (H-K)-(H-K)_{0},
\end{equation}
\begin{equation}
(H-K)_{0} = (H-K)_* - 2.5\,{\rm log}\left(\frac{1+r_H}{1+r_K}\right)
\end{equation}
 where we have indicated with $(H-K)_0$ the dereddened IR color of the source.
From this expression, the visual extinction has been evaluated 
adopting $A_{\rm V} = 15.9\,E(H-K)$ given by the standard interstellar reddening law
 (Rieke \& Lebofsky 1985). The derived values are 
listed in Table 7 and range from \Av=10 mag in IRS3 to \Av=45 mag in 
IRS5a. Our derived values for HH100 IR, IRS2 and IRS5 compare rather well with 
the extinction derived in these sources from the optical depth of the 10\um\, 
silicate and 3\um\, water absorption features, i.e. $\sim$ 25-30 mag for HH100 IR,
 27 mag for IRS2 and 45 mag for IRS5(a+b) (Chen \& Graham 1993, Whittet et al. 1996).
 The low extinction value ($\sim$10 mag) derived for IRS3 is consistent 
with values found towards field stars obscured by the R CrA cloud 
(Whittet et al. 1996).  
 From the adopted spectral type and from the \KK\, magnitude dereddened by the 
derived \Av\, value, we are now 
able to derive the absolute stellar \KK\, magnitude, and thus the intrinsic
stellar luminosity:
\begin{equation}
{\rm log\,L_*/L_{\sun}} = -0.4(M_{bol}-M_{bol,\sun}),
\end{equation}
where
\begin{equation}
M_{bol} = BC + M_K + (V-K)_*
\end{equation}
In this expression, the values for the bolometric correction BC and for
the intrinsic stellar $(V-K)_*$ colors for dwarfs were taken from Kenyon \& Hartmann 
(1995).
Finally, the stellar radius R$_*$ can be also derived given the assumed 
stellar effective temperature. Table 7 lists the L$_*$ and R$_*$ values for 
the five considered sources. Luminosities range between 0.3 and 4.3 L$_{\sun}$,
and
stellar radii between 1.2 and 3.6 R$_{\sun}$, being therefore in the range 
expected for low mass young stars (e.g. Siess et al. 1997).  
Given the stellar luminosity and its effective temperature, the sources can
be positioned in a HR diagram to have an estimate of their mass and age
through the comparison with pre-main sequence evolutionary tracks. This is 
shown in Figure \ref{fig:HR}, where we have considered 
the D'Antona \& Mazzitelli (1997) tracks for stars 
with masses between 0.1 and 2 M$_{\sun}$. 
From these tracks we derive masses between 0.3 and 1 M$_{\sun}$ and ages 
between 10$^5$ and 10$^6$ yr. In order to check for the sensitivity of different
models used to derive the evolutionary tracks, we have also derived masses and 
ages adopting the Siess et al. (2000) models, which are listed in Table 7.
These last models give slightly higher masses and ages with respect
to the D'Antona \& Mazzitelli tracks, with mass variations which are of about 80\%
at worst. 
The considered standard tracks do not take into 
account the perturbation introduced by a high mass accretion rate on the
evolutionary tracks, an effect which could be not negligible for HH100 IR 
and IRS2. Siess et al. (1997, 1999) have computed evolutionary tracks 
for stars which are still accreting a substantial fraction of their 
mass, evaluating the error introduced in the mass and age 
determination by taking the values expected for standard tracks. They
show that while the determination of the (current) mass is not very 
affected by a high accretion rate, the still accreting stars are 
effectively younger than the standard ones, for ages $\la$1.5 10$^6$ 
yr. Thus, it is likely that HH100 IR and IRS2 are actually younger 
than the age of $\sim$5 10$^5$ and 10$^6$ yr, respectively, that we 
infer from the tracks in Fig. \ref{fig:HR}.

The knowledge of the intrinsic stellar luminosity allows us in principle 
to derive which is the fraction of excess luminosity due to accretion
through a comparison with the source bolometric luminosity. The error 
on this L$_{acc}$ determination is largely dominated by the precision through 
which the L$_{bol}$ value is known. With regard to our considered sources, L$_{bol}$ values
estimated through near IR and IRAS photometric data are given for HH100 IR,
IRS2 and for IRS5 (Wilkings 1986, 1992, see Table 1).  IRS5a is 
brighter and its near IR specrum is steeper than IRS5b, which suggests us 
 that the bolometric luminosity of IRS5a dominates over the binary 
system: we assume a ratio 2:1 to separate the $L_{bol}$ value among the two 
components. For IRS6 and IRS3 measurements of the far IR fluxes are not 
available; however the fact that their IR spectra are not very steep  
allow us to assume that the far IR contribution at the L$_{bol}$ value can be 
negligible. We have therefore given an estimate of the L$_{bol}$ values for 
these sources based on their near and mid IR ground based photometry, and 
adding a correction which assumes that the spectrum scales as the wavelength for 
$\lambda$ larger than that corresponding to the 
last observed photometric flux.
For IRS6 we 
furtherly assumes that the two components {\it a} and {\it b} equally 
contribute at the total bolometric luminosity. 

From the comparison of the derived stellar luminosity with the L$_{bol}$ values,
it appears that only HH100 IR and IRS2 have a significant excess in 
luminosity with respect to L$_*$, which amounts to 80 \% and 65 \% of the L$_{bol}$
value, respectively. 
We attribute such an extra luminosity contribution to the energy released in the
accretion process.
In IRS5a, only $\sim$20 \% of the total luminosity
is due to accretion while in the other two sources our method does not allow us to 
measure any significant accretion luminosity. Finally, 
the mass accretion rate can be estimated from the accretion luminosity.
In the hypothesis that
accretion occurs through the circumstellar disk, the total disk accretion
luminosity can be written as (Gullbring et al. 1998):
\begin{equation}
L_{acc} \sim G\frac{M_*\dot{M}}{R_*}(1-\frac{R_*}{R_i})
\end{equation}
where $R_i$ is the inner disk truncation radius which is predicted to be
of the order of 3-5$R_*$.
The $\dot{M}$ values for HH100 IR and IRS2 are $\sim$ 2\,10$^{-6}$
and 3\,10$^{-7}$ M$_{\sun}$ yr$^{-1}$, respectively. Such values are larger 
than the average values derived for T Tauri stars by means of UV excess 
measurements (Gullbring et al. 1998), but lower than the values indirectly 
estimated in Class 0 sources through their mass loss phenomena, 
indicating that an evolution in the 
mass accretion rate during the protostellar life is indeed very likely.

\section{Discussion}
The ability to separately estimate the stellar and circumstellar 
properties of embedded YSOs, 
gives the possibility to infer in a better 
detail the evolutionary properties of these sources, 
to analyse their similarities with optically visible T Tauri stars in terms of veiling and
accretion activity and finally to address quantitatively how 
their accretion properties are connected to other related phenomena such
as jets and winds.

\subsection{Evolutionary stage of the observed sources}

The stellar parameters, accretion luminosity and mass accretion rates derived 
from the analysis of the star photospheric features can be used to assess
 the relative evolutionary phase between the observed sources. We have seen 
that only HH100 IR and IRS2 have a significant fraction of their 
luminosity due to accretion, which is also correlated with the presence 
of both extreme values for the \KK\, and \HH\, veilings and emission 
line activity. 
On the other hand, IRS5a, which like HH100 IR
and IRS2 has been classified as a Class I protostar on the basis of its
SED between 2 and 10\um\, (Wilkings et al. 1997) does not show a significant
accretion activity, as testified by the low $L_{acc}/L_{bol}$,
the absence of emission lines and the low veiling. Presumably,
this source is a T Tauri still embedded in its parental envelope or seen through 
a unfavourable line of sight. 
Finally, IRS6a and IRS3 present characteristics of pure reddened 
photospheres where any sign of activity connected to accretion is very low
or absent.
This analysis shows that YSOs chosen to broadly have 
similar IR colors may have very different characteristics in terms
of accretion properties. This finding suggests us to make an a posteriori
look at the distribution of the IR colors of the considered 
sources to search for a finer separation between stars with different 
accretion properties. In particular the protostars HH100 IR
and IRS2, for which more than 50\% of their luminosity is due to accretion, are located,
 in a near IR color-color diagram, in a region
defined by colors (J-H)$\ga$4 and (H-K)$\ga$2.5. In the same region can be
also put the only other so far identified highly accreting protostar, 
i.e. YLW15 of the $\rho$ Ophiucus cloud ((J-H)$>$5, (H-K)=3.8, Greene \& Lada 2002). 
Sources with a low or absent accretion activity (IRS6 and IRS3) seem to be located 
outside this box, with colors (J-H)$<$4 and (H-K)$<$2.5. In this scheme IRS5a 
is a borderline case,  which remain anomalous. We only note that the colors of 
IRS5(a+b) are considerably bluer than those of the other highly accreting 
protostars ((H-K)/(J-H) = 0.85
vs (H-K)/(J-H) $\sim$ 0.6-0.7), indicating that part of its infrared excess
can be due to scattered radiation. In conclusion we suggest that the real accreting
protostars can be located in a color-color diagram region tighter than 
the previously considered locus for Class I sources (e.g. Lada \& 
Adams 1992), a clue which deserves to be further investigated
on a larger sample basis.\\

It is also interesting to note how the different accretion properties 
of the sources correlate with their position on the HR diagram.
HH100 IR is the youngest source of the sample with an age $\la$ 10$^5$ yr, 
while all the others fall roughly on the same isochrone at 
t$\sim$5\,10$^5$-10$^6$ yr. If taken literally, this result indicates 
that sources with nearly the same
age may have very different $L_{acc}$/$L_{bol}$ values, and thus that the 
age is not the only parameter regulating the end 
of the accretion process.  
In particular, it can be noticed  the different estimated mass accretion 
rate of sources (IRS2, IRS5a and IRS6a) which appear to have a similar age
but different masses, ranging from $\sim$ 0.5 to $\sim$1.5 M$_{\odot}$. 
For these sources there is an evidence that the mass accretion rate increases
with the mass. 
Since the sources belong to the same star 
forming core, they should have started to collapse with the same initial 
cloud infall rate, which depends only on the core kinetic temperature. Therefore,
it is not unexpected that to accumulate a different mass in the same
ellapsing time, their mass accretion rate should have followed a different time evolution. 
Evidences for a correlation between mass accretion
and stellar masses have been indeed reported, although with a broad scatter,
for large samples of T Tauri and very low mass objects (Natta et al. 2004).\

On the other hand, we should also remind that some of the considered stars 
have a large photometric variability as it is indeed common in young embedded stars
(e.g. Kaas 1999, Horrobin et al. 1997). Often variability in IR magnitudes
is associated to variability of spectral features, such as HI and CO lines,
whose emission is directly related to accretion (Biscaya et al. 1997,
Nisini et al. 1994). 
Therefore, if the photometric variability of the considered sources 
is related to changes of the mass accretion rates on short timescales,
as it happens, e.g., in FU Ori-like events, 
the possibility that the three sources with the same age may have
a similar {\it time averaged} mass accretion rate, cannot be excluded.

\subsection{Origin of veiling}

The observed sources present very different values of the IR
veiling. The \KK\, veiling spans from almost 0 in IRS3 and IRS6a, to a value
$\sim$6 in HH100 IR. In T Tauri sources, $r_{K}$ values between 0 and 2
are normally estimated (Johns-Krull \& Valenti 2001, Folha \& Emerson 1999).
Higher values, up to 3-4, have been measured in flat-spectrum and Class I
objects (Greene \& Lada 1997, Doppmann et al. 2003). We however do not find
a clear correlation with the class of the objects, at least for the three Class I
sources of our sample which have $r_{K}$ values of 0.5 (IRS5a), 3 (IRS2)
and 6 (HH100) despite their similar spectral index between 2 and 10$\mu$m.
For these three sources is instead more evident the correlation 
with the estimated accretion luminosity.
In T Tauri stars,
the main considered source of IR veiling is 
the emission from the accretion disk, although,
given the low mass accretion 
rates estimated in T Tauri stars, disk models find some difficulties to 
predict the largest observed veiling values (r$_{k} \sim$ 1-2) 
(Folha \& Emerson 1999, Johns-Krull \& Valenti 2001).
The same models have also difficulties to reproduce the large IR veiling
estimated in Class I sources, despite
the mass accretion rate higher than in T Tauri stars (e.g. Greene \& Lada 1996).
An additional source of veiling can be the emission from the inner circumstellar 
dusty envelopes. Greene \& Lada (2002) estimate that the inner infalling 
envelope can account for about half of the \KK\, veiling they measure in the Class 
I object YLW15.
Calvet et al. (1997),
estimate that the thermal emission of envelopes
with mass infall rates ranging from
10$^{-5}$ and 4\,10$^{-6}$ M$_{\sun}$\,yr$^{-1}$,
can account for a 2.2$\mu$m 
veiling ranging between 3 and 9, thus in the range of the values we find
for HH100 IR and IRS2. The same models predict a \HH\, band
veiling $\sim$ 30-50\% smaller than in the \KK\, band, consistent with
our derived values. 
Finally, the small veiling observed in IRS5a, despite its significant 
IR emission excess, can be explained by assuming that its envelope emission
is mainly due to the scattering of the central object emission. In this case, 
as pointed out by Calvet et al. (1997), the envelope flux contains
the same spectral features as the object and thus does not affect the line EWs. 
Recent models for the SEDs of YSOs (Whitney et al. 2003) show that in Class II
sources seen close to 
edge-on, extinction and scattering dominate
at near-IR wavelengths making the SEDs of these sources similar to those of less 
evolved Class I objects. 

\subsection{Correlation with accretion and ejection activity}
Our observations show that HH100 IR and IRS2, which have most of their luminosity
due to accretion, are also those exhibiting a rich emission line spectrum,
with the strogest lines observed in the source with the highest mass accretion
rate (HH100 IR).

Correlations between the accretion luminosity
and the IR hydrogen lines luminosity have been found in T Tauri stars
(Muzerolle et al. 1998). Such a correlation can be naturally explained 
if the origin of HI lines is in the star magnetospheric accretion
region but also if part of their emission come from ionized outflows, which are
ultimately powered by accretion. 
We can check if the highly accreting objects of our sample still 
follow the same correlation $L_{acc}$-$L_{Br\gamma}$ as T Tauri stars. 
Extending such a correlation to highly accreting embedded objects 
would indeed be important to validate the Br$\gamma$ luminosity as an indirect 
quantitative 
tool derive the accretion luminosity of Class I sources. 
The Br$\gamma$ luminosity of HH100 IR and IRS2,
dereddened for the extinction values listed in Table 7, 
are 5.7\,10$^{-3}$ and 7\,10$^{-4}$ L$_{\sun}$, respectively. Following the fit
derived by Muzerolle et al. (1998), such line
luminosities should be consistent with accretion luminosities $\sim$ 40 and
3 $L_{bol}$ respectively. Considering the large errors associated with the 
Muzerolle et al. least-squares fit, our derived $L_{acc}$ values are
consistent with the relationship found for the T Tauri stars, extending it
to about an order of magnitude larger luminosities.
Br$\gamma$ emission is observed also in IRS5a with a luminosity value $\sim$ 
5\,10$^{-5}$ L$_{\sun}$.
According to the relationship by Muzerolle et al.
the corrisponding $L_{acc}$
value should be a fraction of the bolometric luminosity of the order
of 10$^{-1}$L$_{\sun}$, well inside the locus of the T Tauri star and
in agreement with our roughly estimated value.

It is usually assumed that accretion and ejection of matter are
correlated phenomena and that this latter is needed to remove part
of the angular momentum carried by the accreting material. 
It is therefore worthwhile to see if an high accretion luminosity is
always associated with the presence of outflows.
In this respect it is indicative that HH100 IR is the only source of 
our sample having a clear evidence
of outflowing activity, being the identified exciting source of the HH100 
Hebig Haro nebulosity, and also of the  HH objects
HH101 (located in the SW) and HH99 (in the NE) (Hartigan \& Graham 1987).
A molecular bipolar 
outflow is also associated with this object as evidenced by the
ammonia and HCO$^+$ observations of Anglada et al. (1989) and Anderson et al.
(1997), respectively. 
IRS2, however, despite showing a large accretion luminosity value, does
not show any undoubtful tracer of mass ejection.
We have seen that the weak H$_2$ 2.12$\mu$m line observed in emission 
can be due to both the outflow and the disk and no other lines typical
of jets, such as [FeII] lines, are detected in the spectrum. On a larger
scale no optical HH object seems to be associated to this source 
(Hartigan \& Graham 1987). However, the region closer to IRS2 has not been
investigated neither in H$_2$ imaging, to search for molecular jets, nor
in molecular outflows radio tracers, so we cannot on this basis 
exclude that
a large scale flow is associated with this source.

\section{Conclusion}

We have analyzed the low and medium resolution infrared spectra of a sample
of embedded young stellar objects in the R CrA star forming region
with the aim of constraining their accretion properties and evolutionary
stage. The sample includes three Class I sources (HH100 IR, IRS2 and IRS5)
and two sources with NIR excesses in the Wilking et al. (1997) R CrA core
survey (IRS6 and IRS3).
The main results from this analysis are the following:
\begin{itemize}
\item The low resolution spectra revealed the presence of strong lines
in emission only in two, namely HH100 IR and IRS2, out of the six 
observed sources. The most prominent detected features are permitted 
lines from neutral atoms, mostly from HI (Pa$\beta$, Br$\gamma$ and 
lines from the Brackett series at higher-n) but also OI, NaI and CI. These lines
arise in a partially ionized compact and very dense gas,
likely originating either in the star-disk accretion region, or at the base 
of an expanding wind. The same sources where atomic lines are detected
also present the CO 2.3\um\, bandheads in emission, indicating the presence
of large columns of warm molecular gas, probably located in the 
inner sections of an accretion disk. 
In the medium resolution spectra, also a weak H$_2$ 2.12\um\, line is 
detected in HH100 IR and IRS2, which may be originated either in the 
circumstellar disk or in a molecular jet. 
\item 
The medium resolution spectra show plenty of absorbtion
features typical of late-type stellar photospheres. In HH100 IR and IRS2
such features are very weak, indicating a large amount of IR veiling.
We have derived the effective temperature, extinction, stellar luminosity and
radius of the observed sources, through a comparison with standard star spectra
and IR photometric data. This information has been in turn used to infer
the amount of IR veiling, the accretion luminosity and the mass accretion rates
of the objects. 
\item Only in HH100 IR and IRS2 the accretion luminosity
dominates over the total bolometric luminosity ($L_{acc}$/$L_{bol} \sim$0.8 and 0.6
respectively) and the derived mass accretion rates are of the order of 3\,10$^{-6}$
and 5\,10$^{-7}$ M$_\odot$\,yr$^{-1}$ respectively, i.e. higher by an order
of magnitude with respect to the average values derived in T Tauri stars.
In contrast, in IRS5a only $\sim$2\% of the luminosity is due to accretion.  
In general, we found that there is a correlation between the 
accretion luminosity, the IR veiling and the emission line activity of 
the sources.
\item If compared with standard evolutionary tracks for pre-main sequence evolution,
the derived stellar luminosities and effective temperatures indicate that
HH100 IR is the youngest of the sources, with an age $\sim$10$^{5}$ yr. On the
other hand, IRS2, IRS5a and IRS6a have about the same age ($\sim$5\,10$^{5}$-
10$^{6}$ yr) despite the large differences in their accretion properties.
For these sources there is a hint 
that the mass accretion rate, at a given
age, may depends on the accumulated stellar mass, a result not totally unexpected
if one assumes that the mass accretion rate is a function of the 
mass of the envelope. We cannot however exclude that variations on short timescales
of the accretion rates, related to the sources infrared variability,
may be responsible of the derived differences in the $\dot{M}$ values.
\end{itemize}

Our analysis has shown that sources broadly defined as Class I protostars, 
may indeed be very different in terms of accretion activity. This 
indicates how the definition of the evolutionary stage 
of deeply embedded YSOs by means of IR colors needs to be more carefully
refined. 
In particular, on the basis of our results we suggest that the 
highly accreting protostars (i.e. those for which $L_{acc}$/$L_{bol} >$ 
50\%) can be located in a color-color diagram region characterized by 
near IR colors (J-H)$\ga$4 and (H-K)$\ga$2.5, a clue which remains to be 
further investigated on a larger sample basis.
Low resolution IR spectroscopy can give a better diagnostic about the
accretion activity of young embedded sources, evidenced by  the presence
of emission lines, HI recombination lines and CO overtone emission 
in particular. However, we have shown that the combination of low and
high resolution spectroscopy may be effective in constraining both the stellar
and circumstellar properties of even highly veiled objects, allowing to
directly derive the accretion luminosity and mass accretion rates
back to the first 10$^{-5}$ yr of protostellar evolution.  

\appendix
\section{IR features selected for spectral classification}

\begin{figure}
\resizebox{\hsize}{!}{\includegraphics{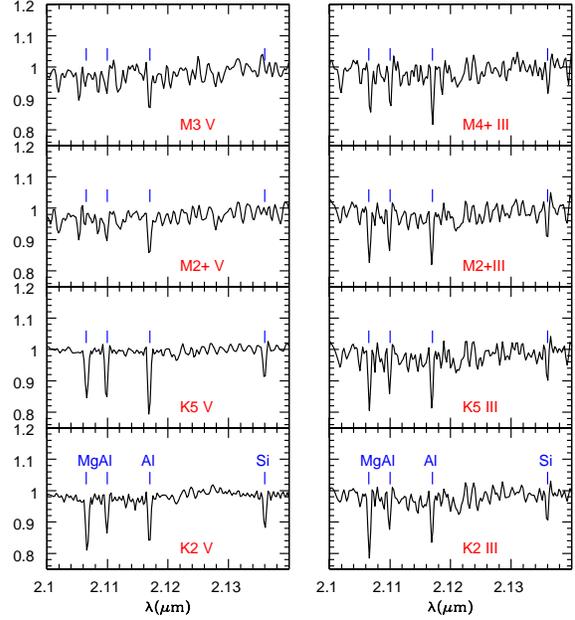}}
   \caption{\label{fig:stand_Al} Spectra of dwarfs (on the left) and giant (on the right) 
   standard stars taken from Wallace \& Hinkle (1997) in a
   spectral segment around 2.1$\mu$m, which includes spectral features 
   of Mg, Al and Si.}
\end{figure}

\begin{figure}
\resizebox{\hsize}{!}{\includegraphics{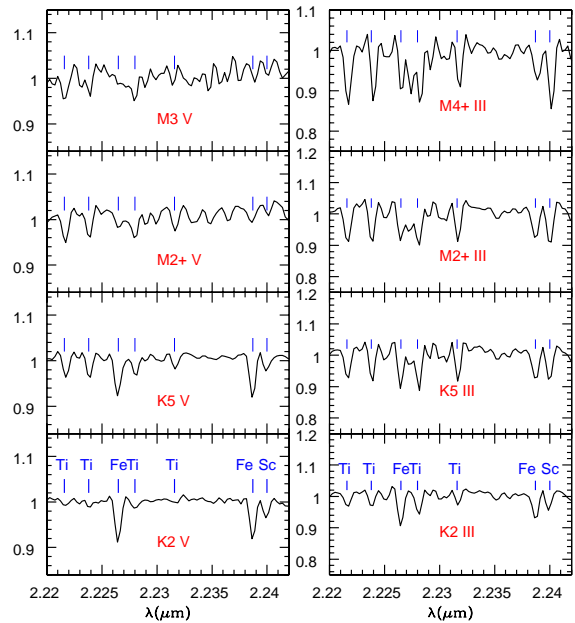}}
   \caption{\label{fig:stand_Ti} Spectra of dwarfs (on the left) and giant (on the right) 
   standard stars taken from Wallace \& Hinkle (1997) in a
   spectral segment around 2.3$\mu$m, which include spectral features of
   Ti, Fe and Sc.}
\end{figure}

In the observed region of the \KK\, band 
we have identified two spectral segments, around 2.12 and 2.23\um, where several  
features can be adopted for spectral classification. 
This can be seen in Figs \ref{fig:stand_Al} and \ref{fig:stand_Ti}, where standard 
spectra of K and M spectral types, 
taken from the Wallace \& Hinkle (1997) catalogue,
are plotted. 
In the spectra 
of dwarfs, around 2.12\um\, Si at 2.1360 \um and Mg at 2.1066 \um\
weaken going from early K to M types and hence their ratios with respect to
the Al lines (2.1099, 2.1169\um), whose EWs remain fairly constant, 
are sensitive to the effective temperature. 

However, in the giant spectra such ratios are less sensitive to 
the effective temperature, 
and therefore a K5V star can be easily confused with a M2III photosphere. 
In order to distinguish also the spectral class, we can use other lines 
around 2.23\um. For example the ratios Ti(2.2217, 2.2240 \um)/Fe(2.2266 \um) and 
Sc(2.2401\um)/Fe(2.2386\um) significantly change 
between dwarfs and giants in K stars (Fig. \ref{fig:stand_Ti}).

For the determination of the spectral type and luminosity class in our
source sample we mainly adopted the above EW ratios in the \KK\, band, and
checked a posteriori that the other observed features were well reproduced.
The use of the features observed in the medium resolution \HH\, band 
spectral segment for the spectral type determination
is limited due to the strong blending affecting most of the lines, and
owing to the fact that the strongest and less blended lines, like Fe I at
1.637$\mu$m and Si I at 1.639$\mu$m, are not very sensitive to the
effective temperature. On the other hand, this fact makes them very useful to
determine the value of the veiling in the \HH\, band once the source 
spectral type have been derived from the \KK\, band features analysis.


\begin{acknowledgements}
We thank Tom Greene for having kindly provided his NIRSPEC spectra of standards. 
We also thank Livia Origlia for useful discussions on spectral classification
and Tom Geballe for helpful comments on the unidentified features. 
This research has made use of NASA's Astrophysics Data System Bibliographic
Services and the SIMBAD database, operated at CDS, Strasbourg, France.
\end{acknowledgements}

\end{document}